\documentstyle[prd,aps]{revtex}
\input epsf
\begin{document}
\title{Trapped and excited $\bbox{w}$ modes of stars with a phase transition and $\bbox {R\geq 5M}$}
\author{Zeferino Andrade}
\address{Department of Physics, University of Utah, Salt Lake City, Utah 84112\\
Department of Physics, University of Wisconsin-Milwaukee, P. O. Box 413, Milwaukee, Wisconsin 53201\footnote{Present address.}}
\maketitle
\begin{abstract}
{The trapped $w$-modes of stars with a first order phase transition (a density
discontinuity) are computed and the excitation of some of the  modes of these stars  by a perturbing shell is investigated. Attention is restricted to odd parity (``axial'') $w$-modes. With $R$ the radius of the star, $M$ its mass, $R_{i}$ the radius of the inner core and $M_{i}$ the mass of such core, it is shown that stars with $R/M\geq 5$ can have several trapped $w$-modes, as long as $R_{i}/M_{i}<2.6$. Excitation of the least damped $w$-mode is confirmed for a few models. All of these stars can only exist however, for values of the ratio between the densities of the two phases, greater than $\sim 46$. We also show that stars with a phase transition and a given value of $R/M$ can have far more trapped modes than a homogeneous single density star with the same value of $R/M$, provided both $R/M$ and $R_{i}/M_{i}$ are smaller than $3$. If the phase transition is very fast, most of the stars with trapped modes are unstable to radial oscillations. We compute the time of instability, and find it to be comparable to the damping of the $w$-mode excited in most cases where $w$-mode excitation is likely. If on the other hand the phase transition is slow, all the stars are stable to radial oscillations.}
\end{abstract}

\section{Introduction and Overview}\label{begining}
The $w$-modes of a fluid star \cite{KOJ,CF,KS1,KBS} represent pulsations of the spacetime geometry of the star, during which the fluid remains almost at rest. In general these modes damp much faster than the fluid modes of the star (although very slowly damped $w$-modes do exist for very compact stars \cite{CF}). This raises the possibility that, in case the $w$-modes are astrophysically excited, they might carry an important part of the gravitational energy radiated by the pulsating star. Although stars of any compactness seem to have $w$-modes \cite{AKK1}, the damping of these modes is so fast for Newtonian stars that even if one of the modes is excited, the star only has time to oscillate a tiny fraction of one period before the vibration dies out. The $w$-modes are thus largely irrelevant compared to the fluid modes for noncompact stars. It is only when the star becomes sufficiently compact that the least damped $w$-modes have damping rates slow enough  that the star might undertake a few complete periods of oscillation at the $w$-mode frequency. 

Recent investigations \cite{AK1,FGB,TSM1,ZP1,ZP2,Ruoff1,TSM2,Ruoff2,FK}
have started to address the conditions under which $w$-modes of compact stars are excited. Nearly all of them have shown that only unrealistically compact stars, with a radius to mass ratio $R/M<3$, have their least-damped $w$-modes strongly excited. For neutron stars (or any other known stars), which have $R/M\geq 5$, the $w$-modes are at best weakly excited. The star pulsates essentially through its fluid modes. It is important however to point out some limitations of these studies.  On the one hand they consider pulsations of static, spherically symmetric stars, i.e.\ they neglect effects due to the rotation of the star. On the other hand, they consider sources of perturbation that probably do not mimic entirely the main aspects of a astrophysically realistic dynamical situation that might leave behind pulsating remnants: the collision of two neutron stars, a supernova explosion or the gravitational collapse of a massive star. There is also the issue of the equation of state used to model the matter of the star. The above studies restricted their attention almost exclusively to polytropic equations of state or equations of state in which the density is a smooth function of the radial position of the matter. While this is reasonable for typical neutron stars, the $w$-modes depend sometimes dramatically on the equation of state and their excitation will likewise be affected. In particular the effect of a first order phase transition (a discontinuity in the density) appears not to have been considered before in connection with $w$-mode studies.

In this work, we use a rather different, even if somewhat unrealistic equation of state, to show that some stars with $R/M\geq 5$ have a $w$-mode spectrum very different from the spectrum of a star described by the usual equations of state. We then go on to show that the least damped of these $w$-modes are excited by matter moving on a spherical shell surrounding the star. As in previous works \cite{ZP1,ZP2} we consider the unperturbed star to be static and we regard the shell as a perturbation of the spherically symmetric spacetime. Because the main intention of this work is to show the effect of the choice of the equation of state on the $w$-modes and on their excitation, we restrict attention to odd parity (axial) perturbations and hence to the odd parity quasi-normal modes of the star, which are all $w$-modes. 

We note that a partial reason why the $w$-modes of stars with $R/M\geq 5$ are barely visible and hence hardly relevant, is because they are very quickly damped. The star simply does not have time to oscillate. On the contrary stars with $R/M<3$ have $w$-modes that are either mildly or, for some equations of state, slowly damped. Usually only for these stars are the $w$-modes significantly excited, carrying sometimes the same or more energy than the excited fluid modes. 

The class of slowly damped modes (sometimes called trapped modes) was first discovered by Chandrasekhar and Ferrari \cite{CF}. They looked at odd parity oscillations of spherically symmetric stars, which are governed by a Schrodinger-like equation. Drawing from studies in atomic physics, they pointed out that a necessary condition for existence of a trapped $w$-mode is that the potential of the Schrodinger equation has a minimum followed by a maximum (a slightly more restrictive necessary condition has been proposed in \cite{ICL}). This is not, however, a sufficient condition: only when the depth and width of the potential well is high or large enough will one (or more) trapped $w$-mode(s) exist. Chandrasekhar and Ferrari then went on to determine the trapped $w$-modes of homogeneous stars and found that they exist only for stars with $2.25<R/M\leq 2.6$. The value $R/M=2.25$ is the highest compactness such stars can reach.

Early on it was believed that in order to have trapped $w$-modes, a star should have values of $R/M$ smaller than $3$. But recent work by Rosquist \cite{ROSQUIST}
and by Karlovini, Rosquist and Samuelsson \cite{KRS1,KRS2,KRS3} suggests that there are equations of state for which stellar models with $R/M\geq 3$ having trapped $w$-modes exist. They studied odd parity oscillations of spherically symmetric stars and showed that certain equations of state lead to a  potential (in the Schrodinger-like equation governing these oscillations) having one or more wells, even when the star has $R/M\geq 3$. If one or more of the wells are deep and broad enough, slowly damped $w$-modes should exist. And if they exist, it is possible that the star might oscillate in one or more of these modes when perturbed by a plausible astrophysical source.

In this work we consider the simplest of the equations of state studied in \cite{ROSQUIST,KRS1,KRS2,KRS3}. The static star has an inner core of constant density $\rho_{+}$, surrounded by an envelope of constant density $\rho_{-}$ such that $\rho_{-}<\rho_{+}$. We will call these double-density stars (DDS). As long as we allow the jump in the density to be sufficiently high, the following argument of Rosquist \cite{ROSQUIST} suggests strongly that some DDS models with arbitrarily high values of $R/M$ might have slowly damped $w$-modes. Let us consider a very compact  star of constant density $\rho_{+}$, with radius $R_{i}$ and mass $M_{i}$ such that $R_{i}/M_{i}\leq 2.6$. This star has trapped $w$-modes. Let us then transform a thin layer of width $\delta$, located at the surface of the star into an envelope with some different but also constant density $\rho_{-}<\rho_{+}$, in such a way that the mass of the whole star is kept approximately constant. The radius of this new star is $R\approx R_{i}(1+3\rho_{+}/\rho_{-}\delta/R_{i})^{1/3}$. It is then clear that choosing $\rho_{+}/\rho_{-}>>R_{i}/\delta>>1$ will lead to $R/M>>R_{i}/M_{i}$, since we kept $M\approx M_{i}$. But we expect the new DDS star to have trapped modes similar to the old star since all we did was add an atmosphere of negligible density to the original star. Additional evidence that this argument should be correct is given in \cite{KRS1}.

It is important to ask what is the physical plausibility of DDS stars (and ultimately of the stars considered in \cite{ROSQUIST,KRS1,KRS2,KRS3}). A realistic neutron star can be seen as being formed by several layers. The density in these layers increases (usually but not always) continuously but over a very short length scale, as we approach the center of the star. It increases from $\rho<10^{6}g/cm^{3}$, in  the surface region to $\rho>10^{15}g/cm^{3}$ in the inner core. A DDS star could then be viewed as modeling one of the abrupt density jumps in a very crude way: by supposing it to be discontinuous. A second case in which a DDS equation of state might be appropriate is during the gravitational collapse of a massive star. There it might be possible that some of the layers experience a jump in the density, eventually well represented by a DDS star. But this dynamical situation would likely be of little relevance to the issue of $w$-mode excitation since the time of collapse would be much smaller than the period of one of the least damped $w$-modes. A third, much more realistic hypothesis, is to regard a DDS as a model of a star with a first order phase transition. Some of these transitions do occur in realistic neutron stars. Directly below the surface of the neutron star, lies the outer crust, a solid region of heavy nuclei in equilibrium with a relativistic degenerate electron gas where the density ranges from $10^{6}g/cm^{3}$ to $4\times 10^{11}g/cm^{3}$. At densities $\rho>10^{7}g/cm^{3}$, the electrons combine with bound nuclear protons to form nuclei with a higher number of neutrons through inverse $\beta$ decay. This neutronization leading from one nuclei to another is a first order phase transition. At $\rho\approx 4\times 10^{11} g/cm^{3}$ neutron drip sets in. The region bellow the outer crust, the inner crust, with densities in the range $4\times 10^{11}g/cm^{3}< \rho < 2\times 10^{14}g/cm^{3}$ is thus a mixture of neutron-rich nuclei together with a superfluid neutron and electron gas. The transition from the solid outer crust to the liquid inner crust can also be seen as a first order phase transition. In both of these phase transitions however, the density jump $\gamma=\rho_{+}/\rho_{-}$ is quite modest, $\approx 3\%$ at the most. Very massive stars might have regions where  $\rho>10^{15}g/cm^{3}$. In those phase transitions from a phase with nucleonic matter to quark matter may occur. Other exotic phase transitions are also predicted at these very high densities: kaon, pion and hyperon condensation. These more speculative phase transitions, might lead to jumps $\gamma$ of up to $10$. For a review about phases in neutron stars and corresponding bibliography from the point of view of nuclear physics see \cite{HHH}.  

Two-phase stars have been modeled both with Newtonian and general relativistic gravity, starting with the pioneering work of Ramsey \cite{RAMSEY} in 1950. He obtained the equilibrium configurations of Newtonian stars with two incompressible layers of constant density and showed that some of these stars might be dynamically unstable to radial oscillations. Ovakimova \cite{OVA} considered the general relativistic analog of Ramsey's star and derived the corresponding stability criteria. Stability criteria for stars whose phases are not of constant density are scarce, except in the case when the core of the denser phase is very small. In this case, Lighthill \cite{LIGHT} generalized Ramsey's stability criteria to two-phase Newtonian stars with phases described by any equation of state, and showed that all such stars are  unstable if the jump in the phase density is bigger than $3/2$ and are stable otherwise. Seidov \cite{ZFS1} extended Lighthill's results to general relativistic two-phase stars and showed that, if the core of the denser phase is very small, general relativity {\em increases} the range of stability of a two-phase star (see also \cite{KAMPFER}). Reviews and recollections by a pioneer in the field of two-phase stars are \cite{ZFS2,ZFS3}. See also \cite{SHZ1,SHZ2,LIND} for several models of equilibrium two-phase stars. 

In this work we will look at a DDS as representing a star with a first order phase transition.  We will also make the additional assumption that each of the layers of the star is incompressible. Since the main point of this work is to show that stars with $R/M\geq 5$ have slowly damped $w$-modes that can be excited, it is important to distinguish between two limiting types of phase transition (a similar distinction is made in \cite{SHZ3}). A slow phase transition is one that happens on a time scale much longer than any of the least damped $w$-modes. As far as $w$-mode excitation, these stars behave as if there were no phase transition at all and the star where made of two different substances in contact along a common boundary. A fast phase transition is one that happens on a timescale much shorter than the period of the least damped $w$-modes. 

Each type of star will behave the same way under odd parity perturbations but differently under even parity perturbations. This is manifest for example if we consider radial oscillations. If the phase transition is very slow then radial oscillations can not exist, the star being incompressible. If on the contrary the transition is fast then the star {\em can} oscillate radially \cite{RAMSEY,OVA,GRIN,BKS}. During this oscillation some of the matter in one of the phases will be converted into the other phase and vice-versa.  
It would appear that which type of star we consider,  is not relevant if our interest is only in odd parity perturbations. However stars with a fast phase transition might in some cases become unstable to the (even parity) radial oscillation, as we mentioned above. If this instability time is much longer than damping time of an excited odd parity $w$-mode, then which type of star we consider is indeed irrelevant. If on the other hand the instability time is comparable (or much smaller) than the damping time of the $w$-mode it is important to make explicit which type of star we are considering. In this later case, it does not make sense to talk of excitation of $w$-modes of a two-phase DDS, while it makes sense to investigate this excitation if the phase transition is slow, for then the star will not be unstable.

We show that stars with $R/M\geq 5$ have trapped modes as long as the compactness of their core, as measured by the ratio of the radius of the core to the mass of the core, $R_{i}/M_{i}$, is such  that $R_{i}/M_{i}<2.6$ and that some of these modes can be excited by the perturbing spherical shell. If the phase transition is slow all of these models are stable (at least to radial oscillations). If the transition is fast, the star can oscillate radially.  We derive an analytic expression for the square, $\omega^{2}$, of the radial frequency of a relativistic DDS (see appendix A). We find that $\omega^{2}<0$ for most (but not all) of the DDS stars whose $w$-modes are excited. The timescale of this instability is in many cases comparable to the period of the excited  $w$-modes.
While homogeneous stars have trapped modes only for $2.25<R/M<2.6$, and the number of these modes decreases as $R/M$ goes from $2.25$ to $2.6$, we will show that DDS models with $R/M<2.6$ and $R_{i}/M_{i}\geq R/M$ can have, for some values of the jump $\gamma$, many more trapped modes than the homogeneous star with the same value of $R/M$ as the DDS.

The paper is organized as follows: The metric coefficients and pressure of a static DDS star and a discussion of the way we parameterize these stars are given in Section \ref{equilibrium}. We then review, in Section \ref{shell}, the perturbation model, a spherical shell surrounding the star, and we present the odd parity equation of motion of the matter in the shell, as well as the specific choice of multipole perturbation we use. In Section \ref{waveq} we present the wave equation governing odd parity perturbations due to the shell and review briefly the method we use to solve this equation. We then investigate in Section \ref{QNmodes} the existence of trapped $w$-modes in DDS stars.
In Section \ref{results}, excitation of $w$-modes by the perturbing shell is considered and the stability of DDS is investigated in Section \ref{stability}. Conclusions are briefly summarized in Section \ref{conclusions}.  The frequency of radial pulsation of a relativistic two-phase star with layers of constant and incompressible density is given in the appendix.  
 
Throughout the paper we use geometric units $G=c=1$ (except where noted) and the conventions of Misner, Thorne and Wheeler\cite{MTW}. 

\section{The background spacetime: Double density homogeneous star}\label{equilibrium}
\subsection{The metric coefficients}
We consider the equilibrium star to be static and spherically symmetric with a radius $R$ and a mass $M$. The star has a inner core of radius $r=R_{i}$, mass $M_{i}$ and constant density $\rho_{+}$, surrounded by an outer layer of density $\rho_{-}<\rho_{+}$. We denote the ratio of the two densities, by $\gamma=\rho_{+}/\rho_{-}>1$. When $\gamma=1$, the outer layer is absent and the star has constant density $\rho_{+}$ throughout its extension. The metric of this background spacetime can be written in the form 
\begin{equation}\label{metric}
ds^{2}=-e^{\nu(r)}dt^{2}+e^{\lambda(r)}dr^{2}+r^{2}[d\theta^{2}+\sin^{2}\theta d\varphi^{2}]\:.
\end{equation}
Outside the star, the metric coefficients are 
\begin{equation}\begin{array}{cc}
e^{\nu(r)}=e^{-\lambda(r)}=1-\frac{2M}{r}\:, & \: r\geq R.
\end{array}
\end{equation}
Inside the star, we need to solve the hydrostatic equilibrium equations of general relativity (see for example \cite{ZP1}, equation (3)) under the assumption of two layers of constant density. This has been done by several authors (see e.g.,\cite{LIND,MVK}). In this work we will use the metric coefficients in
 the form derived in \cite{MVK},
\begin{eqnarray}
& &\label{explambda}
e^{-\lambda(r)}=1-\frac{2m(r)}{r}\:,\\
& &\label{massinner}
m(r)=\left\{
\begin{array}{cc}
\frac{4\pi}{3}\rho_{+}r^{3}\:, & \: r\leq R_{i}\\
\\
\frac{4\pi}{3}\left[\rho_{-}(r^{3}-R^{3}_{i})+\rho_{+}R^{3}_{i}\right]\:, & \: R_{i}\leq r\leq R
\end{array}\right.
\end{eqnarray}
and 
\begin{eqnarray}
& &\label{expnu}
e^{\nu(r)}=\left\{
\begin{array}{cc}
\left[A-B\sqrt{1-\frac{8\pi}{3}\rho_{+}r^{2}}\right]^{2}, & r\leq R_{i}\\
\\
e^{-\lambda(r)}\left[1-4\pi\rho_{-}\sqrt{1-\frac{2M}{R}}\int_{r}^{R}r'e^{3\lambda(r')/2}dr'\right]^{2}, & R_{i}\leq r \leq  R
\end{array}\right.\\
\\
& &
A=\frac{3}{2}(1-\gamma^{-1})\left(1-4\pi\rho_{-}\sqrt{1-\frac{2M}{R}}\int_{R_{i}}^{R}re^{3\lambda(r)/2}dr\right)\sqrt{1-\frac{2M_{i}}{R_{i}}}+\frac{3}{2}\gamma^{-1}\sqrt{1-\frac{2M}{R}}\: ,\\
& &
B=\frac{1}{2}(1-3\gamma^{-1})\left(1-4\pi\rho_{-}\sqrt{1-\frac{2M}{R}}\int_{R_{i}}^{R}re^{3\lambda(r)/2}dr\right)+\frac{3}{2}\gamma^{-1}\sqrt{1-\frac{2M}{R}}\left(1-\frac{2M_{i}}{R_{i}}\right)^{-1/2}\: ,
\end{eqnarray}
where the masses $M_{i}, M$ are (from \ref{massinner})
\begin{equation}
\begin{array}{cc}\label{MiandM}
M_{i}=\frac{4\pi}{3}\rho_{+}R^{3}_{i}\: , & \: M=\frac{4\pi}{3}\left[\rho_{-}(R^{3}-R^{3}_{i})+\rho_{+}R^{3}_{i}\right] \ .
\end{array}
\end{equation}
It is also useful to write down the expression for the pressure inside the star,
\begin{equation}\label{pressure}
p(r)=\left\{\begin{array}{cc}
\rho_{+}\frac{B\sqrt{1-8\pi\rho_{+}r^{2}/3}-A/3}{A-B\sqrt{1-8\pi\rho_{+}r^{2}/3}}, & r\leq R_{i}\\
\\
\rho_{-}\left(-1+\sqrt{1-\frac{2M}{R}}e^{-\nu(r)/2}\right), & R_{i}\leq r\leq R
\end{array}\right. \ .
\end{equation}
\subsection{Specification of a double-density stellar model: Constraints on the values of $R/M$ and $R_{i}/M_{i}$}
We will specify a stellar model by choosing the ratio of its densities, $\gamma$, the compactness of the star, $R/M$, and the compactness of its core, $R_{i}/M_{i}$. It is clear from the expressions (\ref{explambda},\ref{expnu}) for the metric coefficients and (\ref{pressure}) for the pressure, that $R/M$ and $R_{i}/M_{i}$ must both be greater than $2$. However this is not the only constraint on these parameters. Two additional constraints are that $\gamma, R/M, R_{i}/M_{i}$ must be such that $R_{i}/M$ exists and that the pressure (\ref{pressure}) be positive everywhere inside the star.

Lets first address the constraints imposed by the condition that $R_{i}/M$ should exist. To an arbitrary choice of the  parameters $\gamma, R/M, R_{i}/M_{i}$, corresponds three possibilities: a) There is no value of $R_{i}/M$; b) There is one value of $R_{i}/M$; c) There are two values of $R_{i}/M$. To understand this point, we notice that once $\gamma,R/M, R_{i}/M_{i}$ have been specified, $R_{i}/M$ can be determined from (\ref{MiandM}),
\begin{equation}\label{cubic}
\left(\frac{R_{i}}{M}\right)^{3}-\frac{R_{i}}{M_{i}}\frac{\gamma}{\gamma-1}\left(\frac{R_{i}}{M}\right)^{2}+\frac{1}{\gamma-1}\left(\frac{R}{M}\right)^{3}=0\: .
\end{equation}
This cubic equation will have in general three distinct roots. From the theory of cubic equations we can show however that, if
\begin{equation}\label{proib}
\left(\frac{R}{M}\frac{M_{i}}{R_{i}}\right)^{3}>\frac{4}{27}\frac{\gamma^{3}}{(\gamma-1)^{2}}\: ,
\end{equation}
two of the roots are complex and the third is negative. The only relevant set of parameters, $\gamma, R/M, R_{i}/M_{i}$, is then the one for which the inequality (\ref{proib}) does not hold. For that set, there will be three real roots. One of them is always negative and can thus be discarded. The other two, although positive, will only be mathematically admissible if $R_{i}/M\leq R_{i}/M_{i}$ (since $M\geq M_{i}$). From an analysis of the two positive solutions we 
arrive at the following conclusions,
\begin{itemize}
\item If $\gamma<3$, there is always a single solution $R_{i}/M$, if 
\begin{equation}\label{lesscomp}
\frac{R_{i}}{M_{i}}\geq \frac{R}{M}\: .
\end{equation}
This condition states that the core should be less or as compact as the star. There is no solution otherwise. For given $\gamma, R/M$, we have that $R_{i}/M_{i}=R/M$, corresponds to $R=R_{i}, M=M_{i}$, i.e. the stellar model has a single phase with homogeneous density $\rho_{+}$. As we then increase $R_{i}/M_{i}$ above $R/M$ we get ever smaller values of $R_{i}/M$ which, in the limit $R_{i}/M_{i}\rightarrow \infty$, reaches zero.
\item If $\gamma\geq 3$, there is again a single solution $R_{i}/M$, if 
\begin{equation}
\frac{R_{i}}{M_{i}}> \frac{R}{M}\: .
\end{equation}
Otherwise, i.e. if the core is more or as compact as the star,
\begin{equation}
\frac{R_{i}}{M_{i}}\leq \frac{R}{M}\: ,
\end{equation}
and if (\ref{proib}) does not hold, there will always be two solutions $R_{i}/M$. In particular if $R_{i}/M_{i}=R/M$, one of the solutions is the single phase star of density $\rho_{+}$ as before (for which $R=R_{i},M=M_{i}$), but a second solution generally exists with two phases and for which necessarily $R\neq R_{i}, M\neq M_{i}$. A point worth of mention is that for fixed $\gamma, R/M$, one of the two solutions $R_{i}/M$ approaches zero as $\gamma\rightarrow \infty$, while the other approaches $R_{i}/M_{i}$ in the same limit. This is visible in figure 1, for the special case $R/M=5, R_{i}/M_{i}=2.5$.
\end{itemize}
All the above considerations were derived solely from an analysis of the equation (\ref{cubic}). There might be cases however for which a stellar model is apparently possible from the above considerations but it is actually physically impossible because its pressure (\ref{pressure}) is negative in some region of the star \cite{NEGPRESS}. A necessary condition for positivity of the pressure everywhere inside the star, can be obtained by requiring that the pressure at $r=R_{i}$ and at $r=0$, be positive. This leads to the inequality,
\begin{equation}
1<\frac{A}{B}\leq 3\sqrt{1-\frac{2M_{i}}{R_{i}}}\: ,
\end{equation}
which is equivalent to 
\begin{equation}
1-4\pi\rho_{-}\sqrt{1-\frac{2M}{R}}\int_{R_{i}}^{R}re^{3\lambda(r)/2}dr> 0 \: .
\end{equation}
We can use either one of these inequalities to verify whether a set of apparently admissible parameters $\gamma, R/M, R_{i}/M_{i}, R_{i}/M$, characterizes a stellar model with positive pressure everywhere inside the star. If it does not, we discard that set of parameters. Even if these inequalities are valid however, there might be a small set of models for which the pressure is negative. Thus the best way to check whether a stellar model is possible is simply to plot the pressure at all points and see whether or not it is always positive. 

Figure 1 summarizes the results of this section. It shows the physical solutions $R_{i}/M$ of (\ref{cubic}) as a function of $\gamma$, for fixed values of $R/M, R_{i}/M_{i}$. The lower curve, corresponding to the case $R/M=2.5, R_{i}/M_{i}=5$, shows that only one root exists for any $\gamma>1$. As $\gamma\rightarrow \infty$, the root approaches zero, i.e. the larger the jump in density the smaller the core is. If we switch the value of $R/M$ with the value of $R_{i}/M_{i}$, making $R/M>R_{i}/M_{i}$, two roots are possible whenever $\gamma\geq 51.94072268$ (none exists otherwise). The two roots are equal to $R_{i}/M=1.6994$ at this value of $\gamma$, but one of them approaches zero as $\gamma\rightarrow \infty$ (i.e. in this limit it approaches a single layer homogeneous star with density $\rho_{-}$), while the other root approaches $R_{i}/M_{i}$ (i.e. it approaches a single layer homogeneous star with density $\rho_{+}$). These limits can easily be derived from (\ref{cubic}): in the limit $\gamma\rightarrow \infty$, it reduces to a quadratic equation, with the solutions $0$ and $R_{i}/M_{i}$. The behavior of $R_{i}/M$ as a function of $\gamma$ for other given values of $R/M, R_{i}/M_{i}$ is qualitatively the same as the behavior displayed in figure 1.

We would like to stress the following general property of DDS stars with $R/M>R_{i}/M_{i}$. Physical models of these stars only exist for $\gamma\geq \gamma_{min}$. The value $\gamma_{min}$ depends on the relative value of $R/M$ and $R_{i}/M_{i}$. The bigger the difference between these two parameters, the bigger $\gamma$ is.  For example for $R/M=5$ and $R_{i}/M_{i}=2.5$ it is $\gamma_{min}\approx 51.94072268$, while for $R_{i}/M_{i}=2.3$ and $R/M=5$ it is $\gamma_{min}\approx 67.3019402$, and for $R_{i}/M_{i}=2.3$ and $R/M=6$ it is $\gamma_{min}=117.80657$.   
\section{Model: Perturbation by matter moving on a spherical shell}\label{shell}
The source driving the star away from equilibrium is a spherical thin shell of radius $r=R_{shell}>R$, surrounding the star. Most of the details of this model have been presented in \cite{ZP2} for the even parity case. We give in this section only those details that are relevant to odd parity perturbations. 

The stress-energy tensor of the shell is
\begin{equation}
\delta T^{\alpha\beta}_{shell}=\sqrt{1-2M/r}S^{\alpha\beta}\delta[r-R_{shell}]\ .
\end{equation}
We choose the matter in the shell, so that the only non-zero components of the surface stress-energy tensor, $S^{\alpha\beta}$, have odd parity. The metric of the spacetime under this odd parity perturbation can then be written, to first order in the perturbation, as
\begin{equation}\label{pertmetric}
g_{\alpha\beta}=g^{(0)}_{\alpha\beta}+\epsilon h_{\alpha\beta}\ ,
\end{equation}
where the ``(0)'' index denotes the background solution and $\epsilon$ is the perturbation parameter. Because of the spherical symmetry of the shell, we can decompose the components $S_{0\varphi}, S_{\theta\varphi}$, of the surface stress-energy of the shell, in odd parity vector and tensor spherical harmonics,
\begin{mathletters}\label{decomposition}
\begin{eqnarray}
& &
S_{0\varphi}=\sum_{l}S_{l0}^{1}(t)\sin\theta\frac{d Y_{l0}}{d\theta}\ ,\\
& &
S_{\theta\varphi}=\sum_{l}S_{l0}^{2}(t)\frac{1}{2}\left[\sin\theta\frac{d^{2}Y_{l0}}{d\theta^{2}}-\cos\theta\frac{d Y_{l0}}{d\theta}\right]\ .
\end{eqnarray}
\end{mathletters}
Here $Y_{l0}$ is the scalar spherical harmonic. Upon substitution of (\ref{decomposition}) in the equations of motion of the shell (equations (8) of \cite{ZP2}) we get the single equation,
\begin{equation}
\frac{R_{shell}^{2}}{1-2M/R_{shell}}\frac{dS^{1}_{l0}}{dt}+\frac{(l-1)(l+2)}{2}S^{2}_{l0}=0\ .
\end{equation}
By choosing $S^{1}_{l0}(t)$ we uniquely determine $S^{2}_{l0}$. In this work, we make the choice
\begin{equation}\label{gaussian}
S^{1}_{l0}(t)=\epsilon \frac{e^{-at^{2}}}{M}\ .
\end{equation}
Since all perturbation equations will be proportional to $\epsilon$ we will omit it henceforth.

\section{The wave equation governing odd parity perturbations}\label{waveq}
For odd parity perturbations, all physical information can be obtained from a single function $Q_{l0}(r,t)$, that obeys the linear wave equation (see \cite{ZP1} for details),
\begin{equation}\label{oddequation}
\frac{\partial^{2}Q_{l0}}{\partial t^{^2}}-\frac{\partial^{2}Q_{l0}}{\partial r^{*2}}+\left\{l(l+1)-\frac{6m(r)}{r}+4\pi r^{2}[\rho(r)-p(r)]\right\}\frac{Q_{l0}}{r^{2}}e^{\nu(r)}={\cal S}_{l0}(r,t)\ ,
\end{equation}
where $r^{*}$ is the usual {\it tortoise} coordinate,
\begin{equation}
\frac{dr}{dr^{*}}=e^{[\nu(r)-\lambda(r)]/2},
\end{equation}
and the metric coefficients, pressure and remaining quantities characterizing the equilibrium DDS star where given in Section \ref{equilibrium}.
The source term, constructed from the stress energy of the shell, has the form
\begin{eqnarray}
& &
{\cal S}_{l0}(t)=\frac{2D_{l0}}{r^{2}}\left(1-\frac{2M}{r}\right)\left(1-\frac{3M}{r}\right)-\frac{D_{l0,r}}{r}\left(1-\frac{2M}{r}\right)^{2}\ ,\\
& &
D_{l0}=8\pi\sqrt{1-\frac{2M}{R_{shell}}}S^{2}_{l0}(t)\delta[r-R_{shell}]\ .
\end{eqnarray}

We solve equation (\ref{oddequation}), by writing all time dependent quantities as Fourier integrals and imposing, at radial infinity, the condition that the wave be purely outgoing. At $r^{*}\rightarrow \infty$, we can then write,
\begin{equation}\label{waveform}
Q_{l0}(u\equiv t-r^{*})=\frac{1}{2\pi}\int_{-\infty}^{\infty}A_{l0}(\omega)e^{-i\omega u}d\omega \ ,
\end{equation}
where 
\begin{equation}
A_{l0}(\omega)=-\frac{16\pi^{3/2}i\omega}{M\sqrt{a}W_{l}(\omega)[l(l+1)-2]}\sqrt{1-\frac{2M}{R_{shell}}}e^{-\omega^{2}/(4a)}\left[y^{reg}(R_{shell},\omega)+R_{shell}\frac{dy^{reg}}{dr}(R_{shell},\omega)\right].
\end{equation}
Here $y^{reg}$ is the regular solution of equation (\ref{oddequation}) when the source term is zero and $W_{l}(\omega)$ is the Wronskian of the regular and outgoing solutions of (\ref{oddequation}). Details are given in \cite{ZP1}. We would like to point out that despite the fact that the density is discontinuous both at the inner core radius $R_{i}$ and at the surface $r=R$, $Q_{l0}$ and $dQ_{l0}/dr$ should be continuous at both of these points. This follows from the requirement that the intrinsic and extrinsic geometries of the (odd parity) perturbed surfaces of the spacetime  be continuous everywhere \cite{MONCRIEF}. 

 In the remaining sections dealing with odd parity quasinormal modes, we will restrict attention to quadrupole ($\ell=2$) $w$-modes.
\section{The trapped odd parity modes of a double density star}\label{QNmodes}
As is well  known from the work of Chandrasekhar and Ferrari \cite{CF}, a homogeneous star has trapped odd $w$-modes whenever $R/M< 2.6$. The existence of these modes depends first on whether the potential, 
\begin{equation}\label{potential}
V_{l}(r)=e^{\nu(r)}\left\{\frac{l(l+1)}{r^{2}}-\frac{6m(r)}{r^{3}}+4\pi(\rho-p)\right\}\ ,
\end{equation}
of equation (\ref{oddequation}), has a well, and second on how broad and deep the well is. Homogeneous stars with $R>2.6M$ do not have a well and as such they have no trapped $w$-modes. 

In this section, we investigate the effect, on the trapped modes, of adding a second layer of constant density to an already homogeneous star. We have two reasons for undertaking this study. On one hand we want to confirm the argument given in Section \ref{begining}, that DDS stars with $R/M\geq 5$, do have trapped $w$-modes. As we show in the next section, some of these modes are excited by the perturbing shell. On the other hand we want to show that a DDS can have a very different $w$-mode spectrum from a single density layer star, depending on the values of $R/M$ and $R_{i}/M_{i}$. 

To compute the quasi-normal modes we have to solve the equation (\ref{oddequation}) with a zero source term, a complex frequency and the boundary conditions of regularity at the center of the star and a purely outgoing wave at radial infinity. Only for a discrete set of complex frequencies (the odd parity quasi-normal modes of the star) will a solution to this boundary value problem exist. The task of finding these frequencies is usually a complicated numerical problem. However, since we are only interested in the trapped $w$-modes and these are slowly damped, we can use the simpler resonance method \cite{THORNE,CF2,KAK} to compute the frequencies. In this method, we integrate the homogeneous equation (\ref{oddequation}), for a given real frequency, starting at the center of the star and up to a point far away from the star. We know that the general solution of (\ref{oddequation}) at that far away point, is a linear superposition of an ingoing and an outgoing wave. At a quasi-normal frequency, the coefficient of the ingoing solution should drop abruptly to zero. By plotting say, the square of this coefficient, $|A_{ing}|^{2}$, versus the frequency, we can obtain, from the location of the minima, the real part, $\omega_{R}$, of the frequency of  the least damped modes. Since near the minima, $|A_{ing}|^{2}=const[(\omega-\omega_{R})^{2}+\omega_{I}^{2}]$, we can then compute its imaginary part $\omega_{I}$. The method will only work if the imaginary part of the frequency is much smaller than the real part.

Numerical experimentation suggests that a DDS star will have no trapped modes if 1) $R/M>R_{i}/M_{i}>3$ or if 2) $R_{i}/M_{i}>R/M>3$. And the reason is exactly the same as for a single layer homogeneous star: The potential (\ref{potential}) has no well in this case. We then restrict attention, in this section, to DDS models for which either 1) or 2) is not valid. Clearly an exhaustive presentation of results, even in the chosen range of $R/M, R_{i}/M_{i}$ values  is impossible. Instead we will present results for DDS with specific values of $R/M, R_{i}/M_{i}$ and $\gamma$, that we believe illustrate most of the qualitative differences, as well as some similarities, between the trapped $w$-mode spectrum of a DDS and the spectrum of a single density star. 

\subsection{Trapped modes of DDS stars with $\bbox{R/M\geq 5}$}
We turn our attention to the case that is one of the main motivations of this work: Do DDS stars with $R/M\geq 5$ have trapped $w$-modes? A numerical investigation quickly shows that they will but only if $R_{i}/M_{i}<2.6$. This means that all cases of interest for which $R/M\geq 5$, must have a core more compact than the star, i.e. $R_{i}/M_{i}<R/M$. As we recall from Section \ref{equilibrium} (and figure 1), there will then be two $R_{i}/M$ solutions of (\ref{cubic}) as long as $\gamma\geq\gamma_{min}$. None exists otherwise. These two solutions are equal for $\gamma_{min}$, but as $\gamma\rightarrow \infty$, one, that we will call solution A, approaches $R_{i}/M_{i}$ and the other, solution B, approaches 0. As we noted in the final paragraph of Section \ref{equilibrium}, $\gamma_{min}$ will depend on the relative value of $R/M$ and $R_{i}/M_{i}$. It is higher the bigger is the difference between $R/M$ and $R_{i}/M_{i}$. An important conclusion we can draw from all these considerations is that {\it DDS stars with $R/M\geq 5$ and $R_{i}/M_{i}\leq 2.6$, require values of $\gamma>46$}. 

To illustrate the existence of trapped $w$-modes for stars with $R/M\geq 5$, we compare in figure 2, three stellar models: A single density model with $R^{*}/M^{*}=2.3$ and the two DDS models with $R/M=5, R_{i}/M_{i}=2.3, \gamma=68$, the one we called solution A, with $R_{i}/M=1.6442$, and the one we called solution B with $R_{i}/M=1.4647$. The choice $\gamma=68$ is very close to the minimum value of $\gamma$ ($67.3019402$) for which DDS models with $R/M=5, R_{i}/M_{i}=2.3$ exist and as such the two roots A and B are very close to each other. In figure 2a, we compare the potential (\ref{potential}) for each of these stellar models.  The DDS potentials are nearly equal because the two roots $R_{i}/M$ are very close to each other. We thus expect the DDS models to have a similar trapped $w$-mode spectrum, which we confirmed numerically. The DDS potentials are also deeper than the potential of the single layer star and hence we  expect the double layer stars to have more trapped $w$-modes than the single layer star. That this is true is visible in figure 2b, where the least damped of these modes are shown  both for the single layer star and for the DDS with $R_{i}/M=1.4647$. While a single layer star has three or four trapped modes, the DDS star has seven or eight. The values of the frequencies of the two sets of $w$-modes do not appear to be easily related. For reference, we note that the three least damped $w$-modes of the single layer star  have the frequencies  $\omega_{I}M^{*}=0.235+10^{-5}i,\omega_{II}M^{*}=0.316+5.6\times 10^{-4}i$ and $\omega_{III}M^{*}=0.392+7\times 10^{-3}i$. The six least damped frequencies of the DDS model are indicated in figure 2. The jumps that occur at $r=R_{i},R$, in figure 2a, should be expected from (\ref{potential}) and the fact that the density is discontinuous at these points. 
It is also worth noting that while the maximum of the potential lies in the outer layer (of density $\rho_{-}$), the minimum lies deep in the core (with density $\rho_{+}$) of the DDS stars.

What happens to the trapped $w$-mode spectrum of each of the DDS stars with $R/M=5, R_{i}/M_{i}=2.3$, when the jump $\gamma$ is increased and how does these spectra compare to the spectrum of the single layer star with $R^{*}/M^{*}=2.3$?
To understand what happens to the spectrum of solution A, as $\gamma\rightarrow \infty$, we notice that this solution can be obtained from a star with $R^{*}/M^{*}$ and density $\rho_{+}$ by adding to it a layer of density $\rho_{-}$, such as that the resulting DDS star has $R_{i}/M_{i}=R^{*}/M^{*}=2.3$ and $R/M=5$. As $\gamma\rightarrow \infty$, $R_{i}/M\rightarrow R_{i}/M_{i}$ (see figure 1), i.e. in this limit we again obtain the single layer star. Hence we expect that the $w$-mode spectra of solution A, will coincide with the $w$-mode spectra of the homogeneous single layer star for $\gamma\rightarrow \infty$. The two spectra should differ the most when $\gamma$ is near the minimum value allowed (in this case $67.3019402$). The spectrum of $w$-modes of solution B behaves differently in the limit $\gamma\rightarrow \infty$. This solution can be obtained from a star with $R^{*}/M^{*}=5$ and density $\rho_{-}$, by transforming a inner portion into a core of density $\rho_{+}$, in such a way that $R/M=R^{*}/M^{*}=5$ and $R_{i}/M_{i}=2.3$. As $\gamma\rightarrow \infty$, $R_{i}/M\rightarrow 0$, i.e. the inner core shrinks and we approach the original single layer star we started with. But a homogeneous star with $R^{*}/M^{*}=5$ has no trapped $w$-modes (although it has $w$-modes). Therefore we should expect the trapped spectrum of solution B to disappear as $\gamma\rightarrow\infty$. 

To investigate the behavior of the trapped $w$-modes for large $\gamma$, we compare in figure 3a, the potentials of the same three stellar models we used in figure 2, except that we consider a higher density jump for the DDS models, $\gamma=200$. As we can see, the potential of the single layer star is very similar to the potential of the DDS solution A (with $R_{i}/M=2.179$) and as such their least damped modes, shown in figure 3b for solution A, are very similar: $\omega_{I}M=0.221+4\times 10^{-6}$ (compare with $0.235+10^{-5}$), $\omega_{II}M=0.299+2\times 10^{-4}i$ (compare with $0.316+5.6\times 10^{-4}i$) and $\omega_{III}M=0.372+3\times 10^{-3}i$ (compare with $0.392+7\times 10^{-3} i$). On the contrary the potential of solution B (with $R_{i}/M=0.607$) is very different from the other potentials and their least $w$-modes (shown in figure 3b) are therefore very different. Although all three models have four trapped $w$-modes with similar imaginary parts of their frequencies, the real part of the frequencies of solution B are significantly higher than the others as is clear from figure 3b. By increasing $\gamma$ well above $200$, we confirmed that the solution B modes have real frequencies that approach infinite values, i.e. they disappear in the limit $\gamma\rightarrow\infty$ and that the spectrum of solution A approaches the spectrum of the homogeneous star with $R^{*}/M^{*}=2.3$ in the same limit, as we predicted above. A comparison of figures 2b and 3b, also shows that the A solution has more trapped modes the smaller $\gamma$ is (in 2b), and the reason why is evident from both figures 2a and 3a and the arguments above. 

From figures 2 and 3 we can conclude that DDS stars with $R/M=5, R_{i}/M_{i}=2.3$, and values of $\gamma$ near the minimum allowed, have more trapped modes than a single density star with $R/M=2.3$. But as $\gamma$ is increased the number of trapped modes of the DDS stars quickly becomes the same as the number of single layer stars. Thus the effect of the second layer on the trapped modes seems to be more pronounced for the smallest possible values of $\gamma$. 

We only considered the case $R/M=5, R_{i}/M_{i}=2.3$, but DDS stellar models with $R_{i}/M_{i}=2.3$ but higher values of $R/M$ have trapped $w$-modes that behave similarly. The only main difference is that the higher $R/M$ is (with $R_{i}/M_{i}$ fixed), the higher the minimum value of $\gamma$ needs to be for the model to be physically possible, as we discussed in Section \ref{equilibrium}. This makes these DDS highly unrealistic as models of neutron stars with first order phase transitions.  

We also investigated models with $R_{i}/M_{i}=2.5, R/M=5$ and compared their trapped $w$-modes to the trapped modes of single layer stars with $R/M=2.5$, which only have one such mode. Contrarily to what happens when $R_{i}/M_{i}=2.3$, the DDS star never exhibits more than one trapped $w$-mode, no matter what the value of $\gamma$ is. The frequency of this $w$-mode however, is highly dependent on the value of $\gamma$. This dependence can be understood using the same reasoning we presented above. For $R_{i}/M_{i}>2.6$ and $R/M\geq 5$ the DDS stars seem not to have trapped modes.

\subsection{Trapped modes of DDS stars with $\bbox{R/M\leq R_{i}/M_{i}}$}
While DDS stars with $R/M\geq 5$ can have a very different $w$-mode spectra from homogeneous single layer stars, it is for DDS with $R/M\leq R_{i}/M_{i}$ that the difference between the modes of a two layer star and a one layer star is more striking.

Trapped modes exist only for $R/M<3$, so we choose as illustrative examples the DDS stars with $R/M=2.3, R_{i}/M_{i}=5$ and $R/M=R_{i}/M_{i}=2.5$. The interesting aspect of these stars is that, for relatively low values of $\gamma$ (presumably as small as a few percent) the $w$-mode spectrum can be extremely different from the $w$-mode spectrum of a single phase star with the same value of $R/M$. We should keep in mind that, from Section \ref{equilibrium} (and figure 1), if $R/M<R_{i}/M_{i}$, there is only one solution $R_{i}/M$. This solution approaches zero as $\gamma\rightarrow \infty$.

 We thus consider a single layer star with $R^{*}/M^{*}=2.3$ and density $\rho_{-}$. The three least damped $w$-modes of this star, are visible in figure 2b and were discussed above.  We then transform a inner layer of this star into a core of density $\rho_{+}$ and $R_{i}/M_{i}=5$, in such a way that $R/M=R^{*}/M^{*}=2.3$. As we recall from Section \ref{equilibrium}, there will be in general one solution $R_{i}/M$ of (\ref{cubic}) for any value of $\gamma$. For $R/M=2.3, R_{i}/M_{i}=5$ however, models with $\gamma<17.6$ have negative pressures near the core and are thus not allowed. For $\gamma>17.6$, the value of $R_{i}/M$ is maximum for the smallest allowed value of $\gamma$, and it decreases to zero as $\gamma\rightarrow\infty$ (see figure 1). In other words, the configuration $R/M=2.3, R_{i}/M_{i}=5,\gamma=\infty$, is again the  single layer star with density $\rho_{-}$, that we started with. This means that the $w$-mode spectrum of a $R/M=2.3, R_{i}/M_{i}=5$ star, should differ the most from the spectrum of the $R^{*}/M^{*}=2.3$ star, for values of $\gamma$ near $17.6$. The two spectra should approach each other as $\gamma\rightarrow \infty$. This is shown explicitly in figure 4. 

Figure 4a, shows the potential (\ref{potential}), for three stellar models: The single density star with $R^{*}/M^{*}=2.3$ and two $R/M=2.3, R_{i}/M_{i}=5$ DDS stars, one with $\gamma=18$, the other with $\gamma=28$. Figure 4b allows us to visualize the real part of the trapped modes of these models and thus compare the effect of the additional layer. Figure 4b, clearly shows significant differences in the three spectra of oscillation modes. These differences can be partially understood by looking at the shape and depth of the potential well for each of the three models, in figure 4a. There, we see that the smaller the value of $\gamma$ is, the broader and deeper the potential well is. We thus expect the number of trapped $w$-modes for the model $\gamma=18$, to be higher than for the model with $\gamma=28$, and this in turn to be higher than the number of trapped modes of the single layer star. That is precisely what is observed in figure 4b. For $\gamma=18$, there is a very large number of trapped $w$-modes. For $\gamma=28$, however only a few (about 10) remain and only three or four for the single layer star (which is the limit $\gamma\rightarrow \infty$). We should also note from figure 4a, that while the maximum of the potential always resides outside the star (in contrast with the stars in figures 2a and 2b where it resides in the outer layer), the minimum lies deep in the inner layer for the model $\gamma=18$, but lies at the boundary core-envelope for the model $\gamma=28$. 

Figure 4 focused on a particular case, for which $R/M<R_{i}/M_{i}$. It is clear however, that the reasoning we used can be applied, with qualitatively similar results, to stars with $R/M=2.3$, but higher values of $R_{i}/M_{i}$. The only significant change is that as the core becomes less compact (i.e. as $R_{i}/M_{i}$ increases), the minimum allowed value of $\gamma$, for which a solution of (\ref{cubic}) exists and has a positive pressure everywhere, decreases. While for $R_{i}/M_{i}=5$ it was $\approx 17.6$, for $R_{i}/M_{i}\geq 7$ all values of $\gamma>1$ are allowed. 

Our final example, considers the DDS with $R/M=R_{i}/M_{i}=2.5$. This case is interesting because, depending on the value of the density jump, $\gamma$, there can be two, not just one, DDS stars with different values of $R_{i}/M$. Thus, for any value of $\gamma$, one of the solutions of (\ref{cubic}) is $R_{i}/M=2.5$, i.e. $R_{i}=R$, which is just a single layer star with homogeneous density $\rho_{+}$ throughout. For $\gamma\leq 3$ this is the only solution. For $\gamma>3$, however, there is a second solution with two layers and $R_{i}<R$. This solution has however negative pressure in the inner core, for $3.87015<\gamma<48.374$. For this range only the single layer star solution exists. But in the ranges $3<\gamma<3.87015$ and $\gamma>48.374$, the second solution has positive pressure everywhere and we do have two different $R_{i}/M$ solutions with $R/M=R_{i}/M_{i}=2.5$. For $\gamma=3$, the two solutions are the same.
In the limit $\gamma\rightarrow \infty$, the second solution corresponds to a single layer star with density $\rho_{-}$ and $R/M=2.5$ since $R_{i}/M\rightarrow 0$ (see figure 1). It is then clear that the trapped $w$-mode spectrum of each star, in these two limits, will be the same and equal to the $w$-mode spectrum of a single density star with $R/M=2.5$. But for values of $\gamma$ in between these two extrema, the stars modes are very different. 

In figure 5a we show the potential (\ref{potential}), for stellar models with $R/M=R_{i}/M_{i}=2.5$ but four different values of $\gamma$: $3$ (single layer star with $R_{i}=R$), $3.5$ (a double layer star with $R_{i}/M=2.158$), $49$ (a DDS with $R_{i}/M=0.388$) and $200$ (a DDS with a small core, $R_{i}/M=0.184$). The models are identified in the figure by their $R_{i}/M$ values. In figure 5b we show the least damped $w$-modes for the four models depicted in figure 5a. The striking differences can again be  understood from figure 5a and the discussion above. Thus the DDS model with $\gamma=200$ has a single trapped mode, whose value is nearly equal to the single trapped mode of the single layer star with $\gamma=3$. This is expected since the DDS is very close to be a single layer star.  If we compare the DDS model with $\gamma=3.5$ to the single layer model with $\gamma=3$, we notice that a second trapped $w$-mode appears. And many more appear for $\gamma=49$ (whose potential is the deepest in figure 5a). 

We have also studied DDS models with $R/M=R_{i}/M_{i}=2.7$ and $R/M=R_{i}/M_{i}=2.8$. These are similar to the case $R/M=R_{i}/M_{i}=2.5$, in the sense that there is, besides the single layer solution, a second solution with two layers. We found that for certain values of $\gamma$, this second solution has one trapped mode. This should be contrasted with the single density homogeneous solutions with $R/M=2.7,2.8$, which have no trapped modes. Indeed as was shown in \cite{CF}, these stars have trapped modes only for $R/M<2.6$.

One of the conclusions of this section is that the $w$-mode spectrum of a DDS star can differ greatly from the spectrum of a single density star. The differences are more pronounced in the (physically less realistic) case in which $R/M\leq R_{i}/M_{i}$. Indeed in this case, we can form DDS models with a much higher number of trapped $w$-modes than the single layer star trapped modes as is shown in figures 4b and 5b. 

The other conclusion, is that DDS models with $R/M\geq 5$, have trapped modes as long as $R_{i}/M_{i}<2.6$. This requires $\gamma>46$, which might be hard to attain in the universe. It is interesting to note that the more significant difference between double and single density star's trapped mode spectra happens, for values of $\gamma$ near the minimum allowed. 
As we will show in the next section, some of the $w$-modes of these stars can be excited. 

\section{Excitation of $\bbox{w}$-modes of stars with $\bbox{R/M\geq 5}$}\label{results}
To illustrate $w$-mode excitation, by matter moving in the spherical shell surrounding the star, we will consider three DDS models and one single density model. We choose the models so that all have the same value of core compactness, $R_{i}/M_{i}=2.5$. We also choose in all cases, the radius of the shell to be $R_{shell}=50$ and the Gaussian parameter in (\ref{gaussian}) to be $aM^{2}=0.1$. Figures 6 and 7 show the waveform for each of these four models. The general behavior of the waveform is similar in all four cases: Since the shell is located at $r=50$, in all waveforms there is a sharp pulse at $u\approx -r^{*}(50)\approx -56$ corresponding to the motion of the matter in the shell at $t\approx 0$. Part of this pulse travels inward and is reflected by the star. The reflected pulse, whose form and shape depends on the star, will appear at $u\approx 56$. There is then a small transient period, followed usually by ringing at one (or more) of the $w$-mode frequencies of the star.

In figure 6a we present the waveform for a star with homogeneous density $\rho_{+}$ throughout and $R/M=2.5$. The least damped (odd parity) mode of this star
is $\omega M=0.412+0.022i$ (barely visible in figure 5b) and is clearly excited after $u\approx 93$. 

In figure 6b we consider the DDS star with $R/M=R_{i}/M_{i}=2.5, \gamma=3.5$. As we discussed in (\ref{QNmodes}) there are two solutions for these parameters. One is the single density star considered in figure 6a. The other is a DDS star with $R_{i}/M=2.15831$. This star has two clearly visible trapped $w$-modes in figure 5b. Their frequencies are $\omega_{I} M=0.268+10^{-6}i$ and $\omega_{II}M=0.357+0.003i$. The third $w$-mode (barely visible in figure 5b) has frequency $\omega_{III}M=0.412+0.05i$. In figure 6b the star starts ringing at the frequency $\omega_{III}M$ at $u\approx 130$. At $u\approx 170$ there is an interference between modes $\omega_{III}$ and $\omega_{II}$ and after $u\approx 270$ the star rings at the frequency $\omega_{II}$. The least damped mode does not appear to be excited. 

In figure 7 we consider stars with $R/M\geq 5$. We also choose $\gamma$ to be very close to the minimum value allowed for the values of $R/M, R_{i}/M_{i}$ indicated. For $R/M=5, R_{i}/M_{i}=2.5, \gamma=52$, there are two possible models: $R_{i}/M=1.6666$ and $R_{i}/M=1.732$. In figure 7a we consider the first of these models. Ringing of the least damped mode, with frequency $\omega M=0.482+i0.014$, sets in after $u\approx 100$. In figure 7b it is the DDS model with $R/M=6, R_{i}/M_{i}=2.5, R_{i}/M=1.598,\gamma=92$ that is considered. Ringing at the least damped frequency ($\omega M=0.533+i0.021$) sets in after $u\approx 100$.

Two points are worth noting: 1) As is clear from figure 6, the gravitational wave emitted by a single layer star might be considerably different from the wave emitted by a DDS with the same degree of compactness (in this case both stars in figure 6a and in figure 6b have the same value of $R/M$), since the two stars have very different slowly damped quasi-normal modes. The interesting point is that this is true even for small values of the density jump, namely $\gamma=3.5$. But only for very compact stars (i.e. when both $R/M$ and $R_{i}/M_{i}$ are $<3$) will this likely occur.  2) DDS models with $R/M\geq 5$ can have their $w$-modes excited. But this happens at a price: The core $R_{i}/M_{i}<2.6$, which might be a somewhat realistic assumption for neutron stars, but the jump $\gamma>46$, which seems to be unlikely.

\section{Instability of DDS stars with a fast phase transition}\label{stability}
While DDS stars with $R/M\geq 5$ might have their least damped $w$-mode(s) excited, it is important to ask whether these models are stable over timescales greater than the period of the excited $w$-mode(s). To address this question it is useful to divide the DDS stars into two extreme cases: DDS with a very slow phase transition and DDS with a very fast phase transition. We model the first case as if no phase transition occurs and we model the second case as if the phase transition is instantaneous.  Although we might question the validity of this last assumption, we recall that the constant density assumption behind the DDS models is also unrealistic since it allows infinite sound speeds.  Nevertheless we can obtain useful limits on the stability behavior of stellar models with a phase transition and it is with this in mind that we use the two limits of slow and fast phase transitions. 

The adiabatic index on either side of the interface between the two phases is infinite since the phases are supposed to be incompressible and of homogeneous density. In more realistic models of stars with phase transitions, the two phases will have different and finite adiabatic indices. For those models it might be possible to obtain stability criteria in terms of the values of the adiabatic indices, in analogy with what is done in  single phase stars.
 
No matter what the speed of the phase transition, a nonrotating (no $\ell=1$ perturbation allowed) DDS star is stable to odd parity perturbations. This can be easily shown using the fact that all these perturbations (with $\ell\geq 2$ multipoles)must obey the wave equation (\ref{oddequation}). From (\ref{oddequation}), an ``energy-like'' conservation law for the perturbation can be obtained which implies that the perturbation cannot grow without bound at any $r$, i.e. that the star is stable to odd perturbations. For the original proof see \cite{MONCRIEF}. Also \cite{THORNE2}. 
 
Stars with a very slow phase transition (slow compared to the periods of the relevant $w$-modes) are dynamically stable to radial and probably any even parity perturbation. But stars with a fast phase transition,  might be unstable to even parity radial oscillations as shown in \cite{RAMSEY,LIGHT} (Newtonian models) and in \cite{OVA,ZFS1,KAMPFER} (general relativistic models). Indeed in appendix A, we derive the frequency of such oscillations, and from it we conclude that the star is unstable whenever $\omega^{2}<0$, i.e. whenever,
\begin{eqnarray}
& &\nonumber
\frac{1}{R_{i}\sqrt{1-2M/R}}\eta[(\gamma-1)\eta^{3}+4]+\frac{1-2M/R}{R_{i}(\gamma-1)(1-2M_{i}/R_{i})^{3/2}}\times\\
& &
\times\left[3-2\gamma\frac{\sqrt{1-2M_{i}/R_{i}}}{\sqrt{1-2M/R}}+8\pi\rho_{-}\gamma\sqrt{1-\frac{2M_{i}}{R_{i}}}\int_{R_{i}}^{R}re^{3\lambda(r)/2}dr\right]- 36\pi\rho_{-}(1-\frac{2M}{R})\int_{R_{i}}^{R}dr e^{5\lambda(r)/2}<0,
\end{eqnarray} 
a result obtained by Ovakimova\cite{OVA}. 

For very small cores, (i.e. $\eta\rightarrow 0$ or $R_{i}<<R$) the term in square brackets is dominant over the other two terms, and we conclude that the DDS star will be unstable whenever its density jump is
\begin{equation}
\gamma>\frac{3}{2}\frac{\sqrt{1-2M/R}}{\sqrt{1-2M_{i}/R_{i}}}\frac{1}{1-4\pi\rho_{-}\sqrt{1-2M/R}\int_{R_{i}}^{R}re^{3\lambda/2}dr}\ ,
\end{equation}
as was noted first by Seidov \cite{ZFS1}. In the Newtonian limit ($R/M,R_{i}/M_{i}<<1, \rho_{-}M^{2}<<1$) this reduces to the well known result \cite{RAMSEY,LIGHT}, $\gamma>3/2$.

We will now use (\ref{thefrequency}) to compute the frequency of radial oscillation and to find out when are the stars with a fast phase transition unstable. We chose to display results (in figure 8) as plots of the frequency $sgn(\omega^{2})|\omega M|$ (not $\omega^{2}M^{2}$), where $sgn(\omega^{2})=\pm$ depending on the sign of $\omega^{2}$, versus the density jump, $\gamma$. For each plot we keep $R/M$ and $R_{i}/M_{i}$ constant. In this way, when the star is unstable we have negative numbers in the $y$ axis. The inverse of these numbers multiplied by $2\pi$ gives the instability time of the star. The purpose of figure 8 is to give a rough idea of when stars with a fast phase transition are stable and when are they unstable. Even if the model is unstable however, it might be so on a timescale much longer than any of the $w$-modes that are likely to be excited. Therefore we will compare, for a few stellar models, the instability time with the damping time of the least damped $w$-mode(s) to get an idea of the importance, or not, of this radial instability.

\subsection{The case $\bbox{R/M>R_{i}/M{i}}$}
In this case there are in general two solutions of (\ref{cubic}) (see figure 1)for $\gamma\geq \gamma_{min}$, one with a small core and the other with a big core. We represent in figure 8a, three families of models with fixed values of $R/M, R_{i}/M_{i}$ but different values of $\gamma$. The small core solution is represented by a dashed line, and the big core solution by a solid line. The frequencies are monotonous functions of $\gamma$ for each of the solutions. We see that the small core solution is always unstable for models with $R_{i}/M_{i}<2.6$. The solution with the big core might be stable if $\gamma$ and $R_{i}/M_{i}$ are high enough, (for example almost all models with $R_{i}/M_{i}=2.3$ are unstable, but many of the models in the second solution with $R_{i}/M_{i}=2.5$ and high enough $\gamma$ are stable). In particular all models considered in Section \ref{results} are unstable.
Is the instability relevant in the context of $w$-mode excitation? Let us consider the model $R/M=5, R_{i}/M_{i}=2.5,\gamma=52$ and $R_{i}/M=1.67$ (figure 7a above). The excited $w$ mode has a damping time of $t_{damp}/M=2\pi/0.014\approx 449$. The instability time is $2\pi/0.072\approx 87.3$. For a star with one solar mass, these times are $t_{damp}=2.2 msec, t_{inst}=0.4 msec$. Therefore for DDS stars with these parameters, and a fast phase transition, it is not relevant to talk of $w$-mode excitation. A similar computation for the model in figure 7b leads to the same conclusion: The damping time of the excited $w$-mode is $t_{damp}/M\approx 299 $ and the instability time is $t_{inst}/M=2\pi/0.06\approx 105$. For a star with one solar mass, these times are $t_{damp}=1.5 msec, t_{inst}=0.5 msec$. All models with two layers in figures 2 and 3 are also unstable.

But there are stable models: For $R/M=5, R_{i}/M_{i}=2.6, \gamma=77$, the small core solution ($R_{i}/M=1.005$) is unstable, but the big core solution ($R_{i}/M=2.331$) is stable. In general however, all solutions with $R/M\geq 5$, $R_{i}/M_{i}<2.6$, for which the likelihood of $w$-mode excitation seems the highest, will be unstable for the least possible values of $\gamma$.  
  
\subsection{The case $\bbox{R/M<R_{i}/M_{i}}$}
As we remember from Section \ref{equilibrium}, if $R/M<R_{i}/M_{i}$, there is only one solution of (\ref{cubic}). In figure 8b we show some curves frequency versus $\gamma$ for a few stellar models with quite small value of $R_{i}/M_{i}$. The form of these curves is similar for all stars with $R/M<R_{i}/M_{i}$:  for small $\gamma$ (near $1$), the term in square brackets in (\ref{thefrequency}) completely dominates the other terms and the frequency has its maximum value. This value is positive. As $\gamma$ increases, the frequency drops, becoming negative for some models or staying positive for others, until eventually it reaches a minimum, after which it rises again approaching an asymptotic limit as $\gamma\rightarrow\infty$. In figure 8b we consider three curves corresponding to DDS models with the same value of $R_{i}/M_{i}=5$ but different values of $R/M$: $2.3, 2.5$ and $4$. We also consider a curve corresponding to models  with $R_{i}/M_{i}=2.3$ and $R/M=50$. The three curves  with $R_{i}/M_{i}=5$, show that above a certain value of the jump $\gamma$, all the models are unstable. The frequency decreases as the jump in the density increases meaning that the instability time of these models {\it increases}. For small values of $\gamma$ the instability time is a minimum though. And this minimum can  be much smaller (by several orders of magnitude) or comparable to the damping time of the $w$-modes. The conclusion is that models with $R/M<R_{i}/M_{i}$, having trapped modes will be certainly unstable for almost all small values of  $\gamma$ and that instability certainly kills the star before significant $w$-mode ringing goes on for small values of $\gamma$. However for very high values of $\gamma$, the instability might allow  $w$-mode ringing. 

It is instructive to look at some of the models we considered in Sections \ref{QNmodes} and \ref{results}. The model considered in figure 6a is obviously stable (it is a single density homogeneous star). The model in figure 6b is also stable. The remaining models, all with $R/M=R_{i}/M_{i}=2.5$, in figure 5 are both unstable. All the models in figure 4 (with the exception of the homogeneous model) are unstable, as is evident from figure 8b. 
\section{Conclusions}\label{conclusions}
We have shown in Section \ref{QNmodes} that adding a second layer to a homogeneous density star, can change completely the $w$-mode spectrum. In particular stars with $R/M\geq 5$, very compact cores, $R_{i}/M_{i}<2.6$ and high density jumps $\gamma>46$, can have trapped (very slowly damped ) $w$-modes. Some of these modes are excited as we showed in Section \ref{results}. But where the effect of a second layer is more striking is when both $R/M$ and $R_{i}/M_{i}$ are smaller than $3$. In that case, even for very small values of the density jump ($\gamma\sim 18$ in figure 4 and $\gamma\sim 3$ in figure 5) we might have stars with completely different $w$-mode spectra from the spectra of homogeneous stars. We have found that DDS models with $R/M=2.5$, can have a very large number of trapped $w$-modes. By contrast, homogeneous stars with a very large number of trapped $w$-modes, require values of $R/M\approx 2.25$, very near the maximum compactness of the star. 

Most of the DDS models with trapped modes, the most likely to have their $w$-modes excited, are unstable to radial oscillations, if the phase transition, at the origin of the density jump, is very fast. But if the transition of one phase to the other is slow, the models are stable. It would be of interest to investigate whether, in realistic stars, the phase transition happens on a slow or fast timescale, since this might determine whether such stars can or can not exist.

Although we showed that DDS stars with $R/M\geq 5$ can have their $w$-modes excited, all of these stars require relatively large discontinuities of the density. Values that are unlikely to occur in realistic neutron stars. It is thus worthwhile to use different, more realistic, equations of state with a density discontinuity and see if the phenomena observed in this work for relatively high $\gamma$, occurs for realistic values of $\gamma$.  For example does dropping the incompressibility and/or the constant density hypothesis change in a significant way $w$-mode excitation?

It is also useful to ask whether the results reported here depend on the existence of a density discontinuity. We know that in a realistic neutron star, the density is usually continuous but might change over very small length scales. Thus it would be interesting to consider what would happen if the discontinuity is smoothed out and instead we consider the jump from $\rho_{+}$ to $\rho_{-}$ to happen continuously but over a very short length.  Would it be possible, to have trapped $w$-modes and significant $w$-mode excitation, for stars with continuous density throughout and $R/M\geq 5$? An obvious idea based on this work, is to consider stars with two regions of very different densities (a inner higher density layer, surrounded by an outer low density layer), connected by a thin shell where the density drops abruptly, but continuously, from its high to its low value. Will the $w$-mode spectrum and excitation depend on the detailed form of the transition from high to low density? If yes, what are the relevant factors?

We restricted attention in this work, to odd parity perturbations. It is of some interest to study even parity perturbations of DDS stars and see whether the excitation of $w$-modes of stars with $R/M\geq 5$ and $R_{i}/M_{i}<2.6$ is comparable to the fluid mode excitation. We have started addressing some of these questions, in the hope they might shed some light on the issue of the relevance of $w$-mode excitation in realistic stars.

\section{Acknowledgments}\label{acknow}
I would like to thank Hugh Van Horn for pointing to me a crucial detail at the beginning of this work and Greg Comer and Ben Owen for reading a first draft of this work and making several useful comments and suggestions. In particular, I am deeply indebted to Zakir Seidov for many clarifying discussions and information, including references, about two-phase stars and the associated radial instability.  
Appreciation is expressed to John Friedman and Richard Price for useful comments. This work was partially supported by SFRH/BPD/1618/2000 grant from FCT (Portugal) and by NSF grants PHY-9970821 (Milwaukee) and PHY-9734871 (Utah).
\appendix
\section{Frequency of radial oscillation of a relativistic two-phase star}
We derive the frequency of radial oscillation of a star with two-phases, each of constant density and under the assumption that the fluid is incompressible in each phase. We also assume that the phase transition is so fast that at any instant during the radial vibration, the pressure on either side of the phase boundary is equal to the pressure at which the two phases are in equilibrium. 

Under these assumptions, Grinfeld \cite{GRIN} and Bisnovatyi-Kogan and Seidov \cite{BKS}, obtained the frequency of radial oscillation of a two-phase Newtonian star and from it the stability criteria previously derived by Ramsey \cite{RAMSEY}.  Haensel {\it et al} \cite{SHZ3}, Bisnovatyi-Kogan and Seidov \cite{BKS} and Migdal\cite{MIG} have also computed the radial frequency of oscillation of Newtonian two-phase stars under more realistic assumptions.
The frequency of a relativistic two-phase star appears not to have been derived before, although Ovakimova \cite{OVA} obtained the criteria of instability of these stars using a method that did not require the knowledge of the frequency.
 
In the following, let $\xi$ denote the radial displacement of a fluid element from its equilibrium position and let the quantity $\delta h$, be defined by
\begin{equation}
\delta h=\frac{\delta p}{p+\rho},
\end{equation}
where $\delta p$ is the Eulerian variation of the pressure, and $p,\rho$ are the equilibrium values of the pressure and density. A superscript ``$-$'' means the quantity is evaluated at $r=R_{i}-\epsilon$ and a superscript ``$+$'' means it is evaluated at $r=R_{i}+\epsilon$ ($\epsilon<<1$). It is straightforward to show, under the above assumptions, that the quantities $\delta h$ and $\xi$ must obey the following three boundary conditions:

1. Continuity of $\delta h$ at the unperturbed phase boundary
\begin{equation}\label{c1}
\delta h^{+}(R_{i})=\delta h^{-}(R_{i}).
\end{equation}

2. Continuity of the Lagrangian perturbation of the pressure at $r=R_{i}$,
\begin{equation}
[p(R_{i})+\rho_{+}]\delta h^{-}+\xi^{-}\frac{dp^{-}}{dr}=[p(R_{i})+\rho_{-}]\delta h^{+}+\xi^{+}\frac{dp^{-}}{dr}.
\end{equation}
Using (\ref{c1}) and the equation of hydrostatic equilibrium to eliminate $dp^{\pm}/dr$, this equation reduces to
\begin{equation}\label{c2}
(\rho_{+}-\rho_{-})\delta h=\left\{\xi^{-}(R_{i})[\rho_{+}+p(R_{i})]-\xi^{+}(R_{i})[\rho_{-}+p(R_{i})]\right\}\frac{M_{i}+4\pi R_{i}^{3}p(R_{i})}{R_{i}(R_{i}-2M_{i})}.
\end{equation}

3. Vanishing of the Lagrangian perturbation at the surface of the star
\begin{equation}\label{c3}
\delta h(R)-\xi(R)\frac{M}{R(R-2M)}=0.
\end{equation}
We should note that $\xi^{+}(R_{i})\neq \xi^{-}(R_{i})$, because there is a flux of mass across the phase boundary due to the phase transition. This should be contrasted with other studies \cite{FINN,MCDERMOTT} of oscillations of stars with density discontinuities but without phase transitions, for which $\xi^{+}(R_{i})=\xi^{-}(R_{i})$. 
 
Imposing the three boundary conditions (\ref{c1},\ref{c2},\ref{c3}) to the solutions of the equations of radial oscillations, leads directly to an analytic expression for the frequency of such oscillations, as we now show.  We will follow the equations and notation in chapter 26 of \cite{MTW} in the remainder of this appendix, taking care of the difference between the form of the metric (\ref{metric}) we use here, and the form they use.

The Eulerian perturbation of the density is given by equation (26.11) of \cite{MTW}. Noting that the star is incompressible, this equation reduces to
\begin{equation}\label{p1}
\frac{1}{r^{2}}e^{-\lambda/2}\frac{d}{dr}\left[r^{2}e^{\lambda/2}\xi\right]+\frac{\delta \lambda}{2}=0.
\end{equation}
But from equation (26.15) of \cite{MTW}, we know that
\begin{equation}\label{inter}
\delta \lambda=-(\nu_{,r}+\lambda_{,r})\xi.
\end{equation}
Substituting this equation in (\ref{p1}), we get the following equation for $\xi$,
\begin{equation}
\frac{1}{r^{2}\xi}\frac{d}{dr}[r^{2}\xi]=\frac{\nu_{,r}}{2},
\end{equation}
whose integral is $\xi(r,t)=A(t)e^{\nu(r)/2}/r^{2}$. Since $\xi$ must be regular at the origin we have
\begin{equation}\label{p22}
\xi(r,t)=\left\{\begin{array}{cc}
 0, & r< R_{i}\\
\\
A(t)\frac{e^{\nu(r)/2}}{r^{2}}, & R_{i}<r<R.
\end{array}\right.
\end{equation}
To determine $\delta h$, we need the relativistic Euler equation, (\cite{MTW}, equation (26.18)),
\begin{equation}\label{p3}
e^{\lambda-\nu}\xi_{,tt}=-\delta h_{,r}-\frac{1}{2}\delta \nu_{,r},
\end{equation}
and 
\begin{equation}\label{p4}
\delta \nu_{,r}=\left[\nu_{,r}+\frac{1}{r}\right]\delta\lambda +8\pi r(p+\rho)\delta h e^{\lambda},
\end{equation}
which is obtained by combining equations (26.13b) and (26.14) of \cite{MTW}.

Substituting $\delta \nu_{,r}$ in (\ref{p3}), by its expression in (\ref{p4}), (and using also \ref{inter}) we obtain the following equation for $\delta h$,
\begin{equation}\label{deleq}
e^{\lambda-\nu}\xi_{,tt}=-\delta h_{,r}+\frac{1}{2}(\nu_{,r}+\lambda_{,r})\xi\left[\nu_{,r}+\frac{1}{r}\right]-4\pi r (p+\rho)\delta h e^{\lambda}.
\end{equation}
Let us assume that all quantities have a time dependence of the form $e^{i\omega t}$ where $\omega$ is the frequency of radial oscillation.
Upon substitution of the expression (\ref{p22}) in (\ref{deleq}), we obtain, after performing the radial integration,
\begin{equation}
\delta h e^{[\nu(r)+\lambda(r)]/2}=\left\{
\begin{array}{cc}
B, & r<R_{i}\\ 
\\
A\int_{R_{i}}^{r}dr\left\{\omega^{2}e^{3\lambda/2}\frac{1}{r^{2}}+\frac{1}{2}(\nu_{,r}+\lambda_{,r})\frac{1}{r^{2}}e^{\nu+\lambda/2}\left[\nu_{,r}+\frac{1}{r}\right]\right\}+C, & R_{i}<r<R .
\end{array}\right.
\end{equation}
If we now impose the three boundary conditions (\ref{c1},\ref{c2},\ref{c3})
we can determine $\omega$ directly. The result is
\begin{eqnarray}
& &
\omega^{2}=\frac{1}{\int_{R_{i}}^{R}dr \frac{e^{3\lambda/2}}{r^{2}}}\left\{-\frac{B}{A}+\frac{e^{\nu(R)/2}}{R^{3}}\frac{M}{R-2M}-\frac{1}{2}\int_{R_{i}}^{R}dr(\nu_{,r}+\lambda_{,r})\frac{1}{r^{2}}e^{\nu+\lambda/2}\left[\nu_{,r}+\frac{1}{r}\right]\right\}\\
& &
\frac{B}{A}=-\frac{e^{\nu(R_{i})+3\lambda(R_{i})/2}}{\rho_{+}-\rho_{-}}[p(R_{i})+\rho_{-}]\frac{M_{i}+4\pi R^{3}_{i}p(R_{i})}{R^{4}_{i}}.
\end{eqnarray}
Using the metric coefficients in Section \ref{equilibrium} and the general relativistic equations of hydrostatic equilibrium, we can, through lengthy but straightforward algebra, rewrite the frequency of radial oscillations in the somewhat simpler form,
\begin{eqnarray}
& &\nonumber
\omega^{2}=\frac{4\pi\rho_{-}}{3}\frac{1}{\int_{R_{i}}^{R}dr e^{3\lambda/2}/r^{2}}\left\{\frac{1}{R_{i}\sqrt{1-2M/R}}\eta[(\gamma-1)\eta^{3}+4]+\frac{1-2M/R}{R_{i}(\gamma-1)(1-2M_{i}/R_{i})^{3/2}}\times\right.\\
& &\label{thefrequency}
\left.\times\left[3-2\gamma\frac{\sqrt{1-2M_{i}/R_{i}}}{\sqrt{1-2M/R}}+8\pi\rho_{-}\gamma\sqrt{1-\frac{2M_{i}}{R_{i}}}\int_{R_{i}}^{R}re^{3\lambda(r)/2}dr\right]- 36\pi\rho_{-}(1-\frac{2M}{R})\int_{R_{i}}^{R}dr e^{5\lambda(r)/2}\right\}.
\end{eqnarray}
where $\eta=R_{i}/R$ is the ratio between the radius of the core and the radius of the star. 



\begin{figure}
\hspace*{.1\textwidth}\epsfxsize=0.7\textwidth
\epsfbox{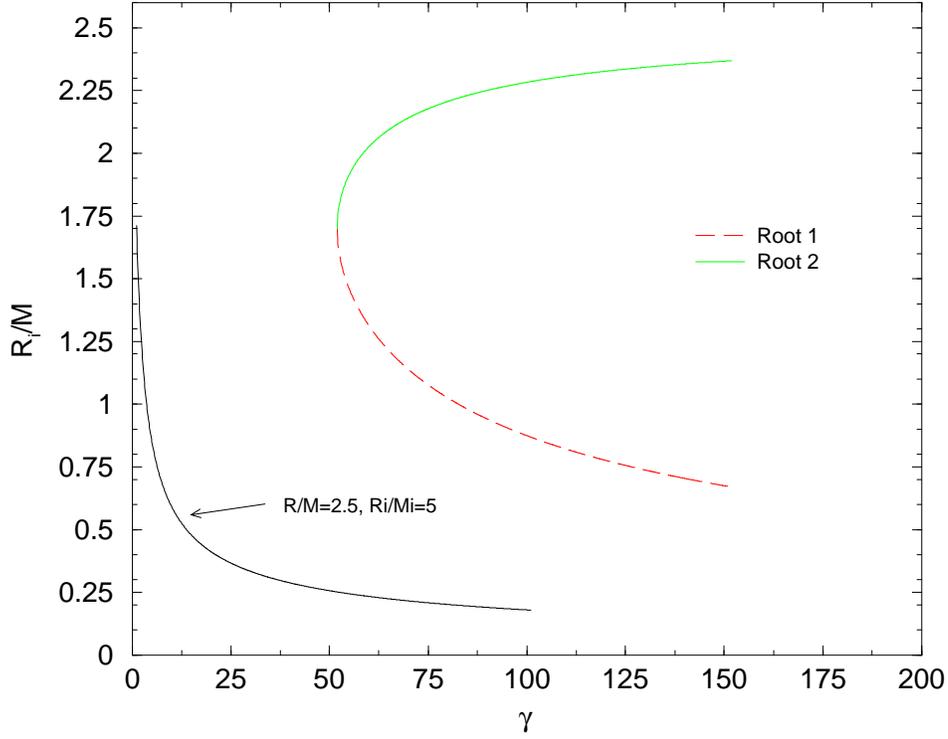}
\caption{Roots $R_{i}/M$ of equation (\ref{cubic}) when $R/M=5, R_{i}/M_{i}=2.5$ or vice-versa, $R/M=2.5, R_{i}/M_{i}=5$, plotted as a function of $\gamma$.}
\end{figure}

\begin{figure}
\hspace*{.1\textwidth}
\epsfxsize=0.35\textwidth
\epsfbox{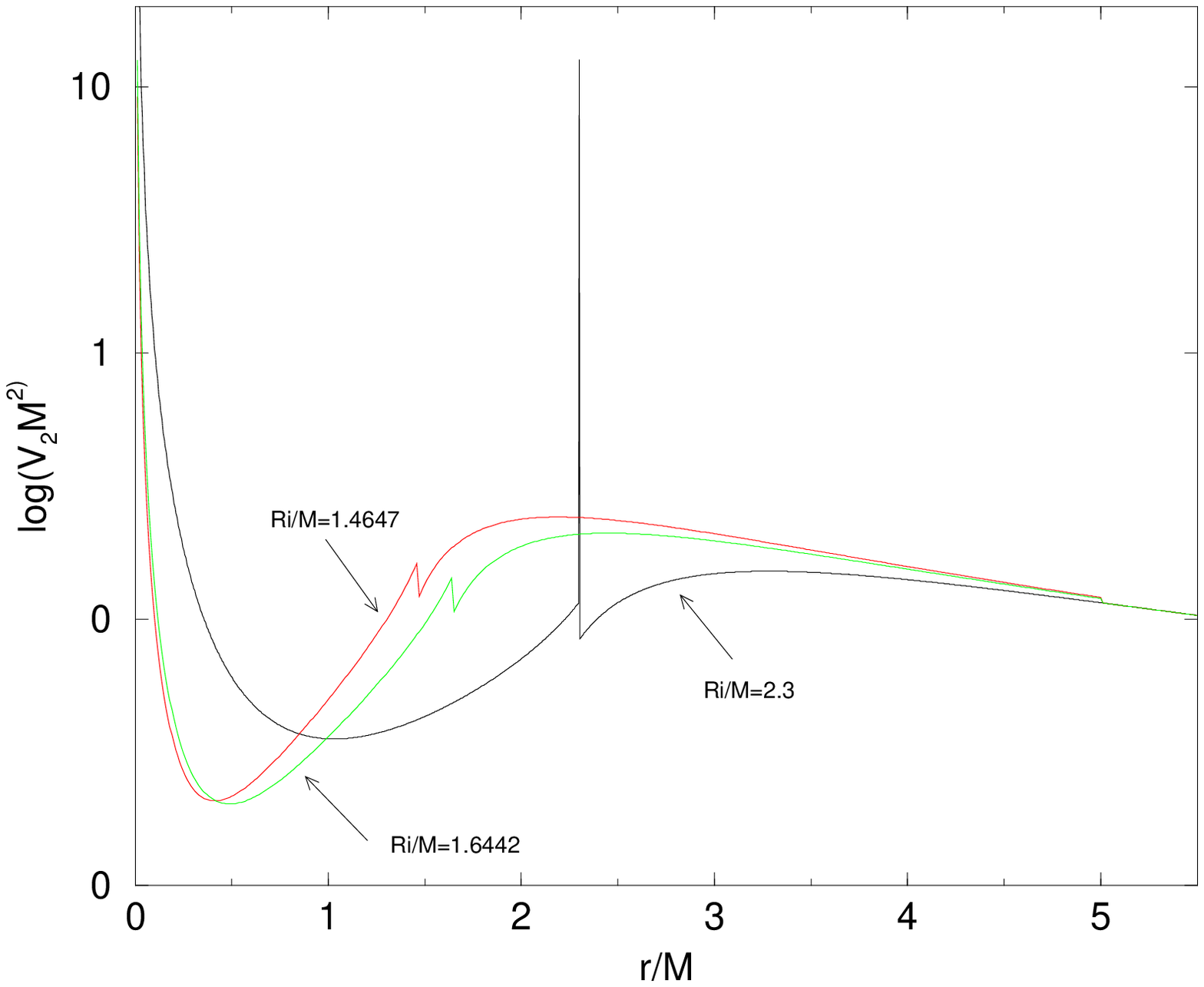}\hspace*{.1\textwidth}\epsfxsize=0.35\textwidth
\epsfbox{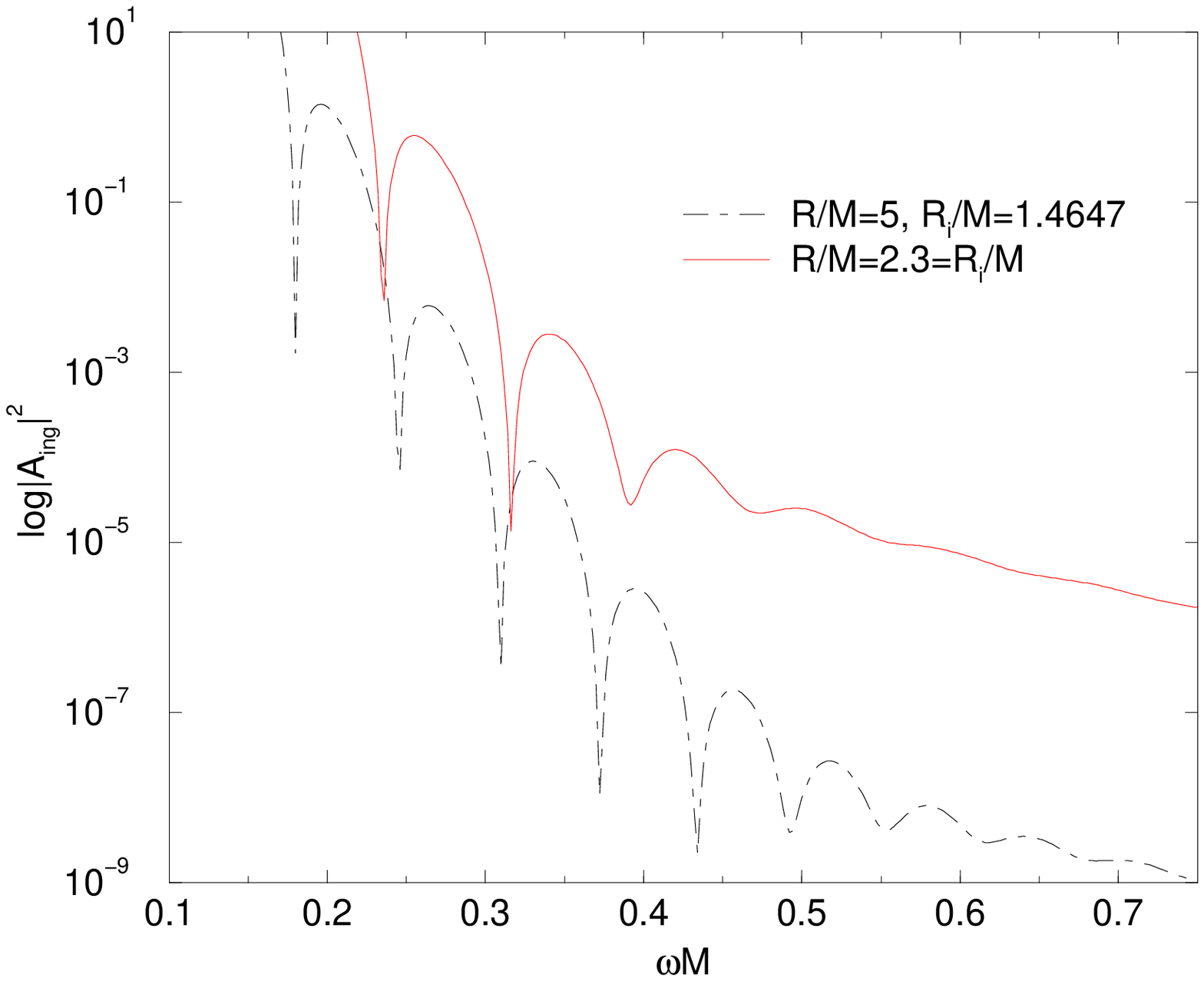}
\caption{Figure 2a displays the odd parity potential for $\ell=2$ and three stellar models: a single density model with $R/M=R_{i}/M_{i}=2.3$ and two double density models both with $\gamma=68, R/M=5, R_{i}/M_{i}=2.3$ but $R_{i}/M=1.4647$ and $R_{i}/M=1.6442$. Figure 2b shows the least damped $w$-modes of the single layer star and the double layer star with $R_{i}/M=1.4647$. The six lowest frequencies of this double star are $\omega_{I}M=0.180+10^{-8}i$, $\omega_{II}M=0.245+10^{-7}i$, $\omega_{III}M=0.309+29\times 10^{-6}i$, $\omega_{IV}M=0.373+15\times 10^{-5}i$, $\omega_{V}M=0.434+10^{-3}i$, $\omega_{VI}M=0.493+5\times 10^{-3}i$.}
\end{figure}

\begin{figure}
\hspace*{.1\textwidth}
\epsfxsize=0.35\textwidth
\epsfbox{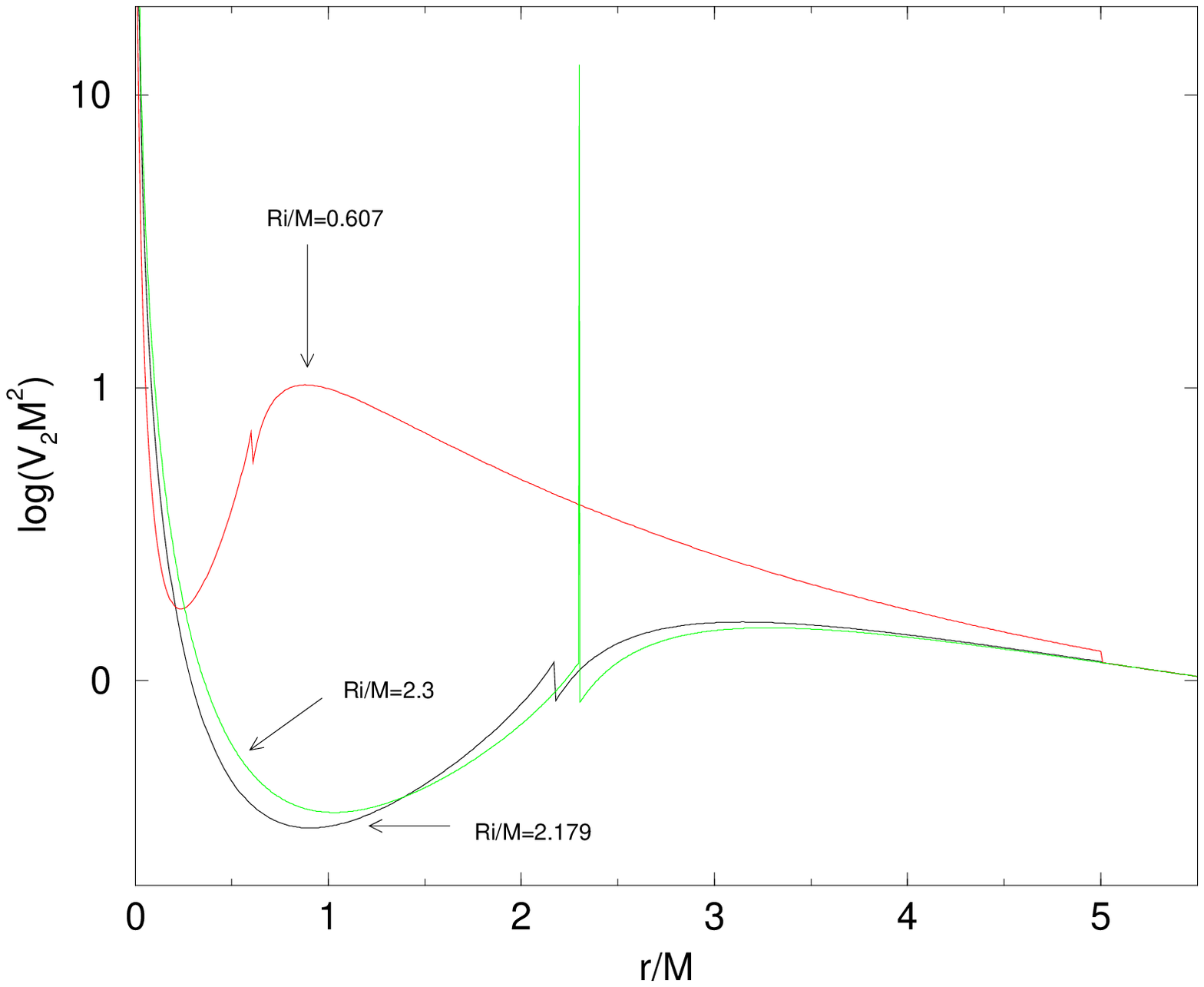}\hspace*{.1\textwidth}\epsfxsize=0.35\textwidth
\epsfbox{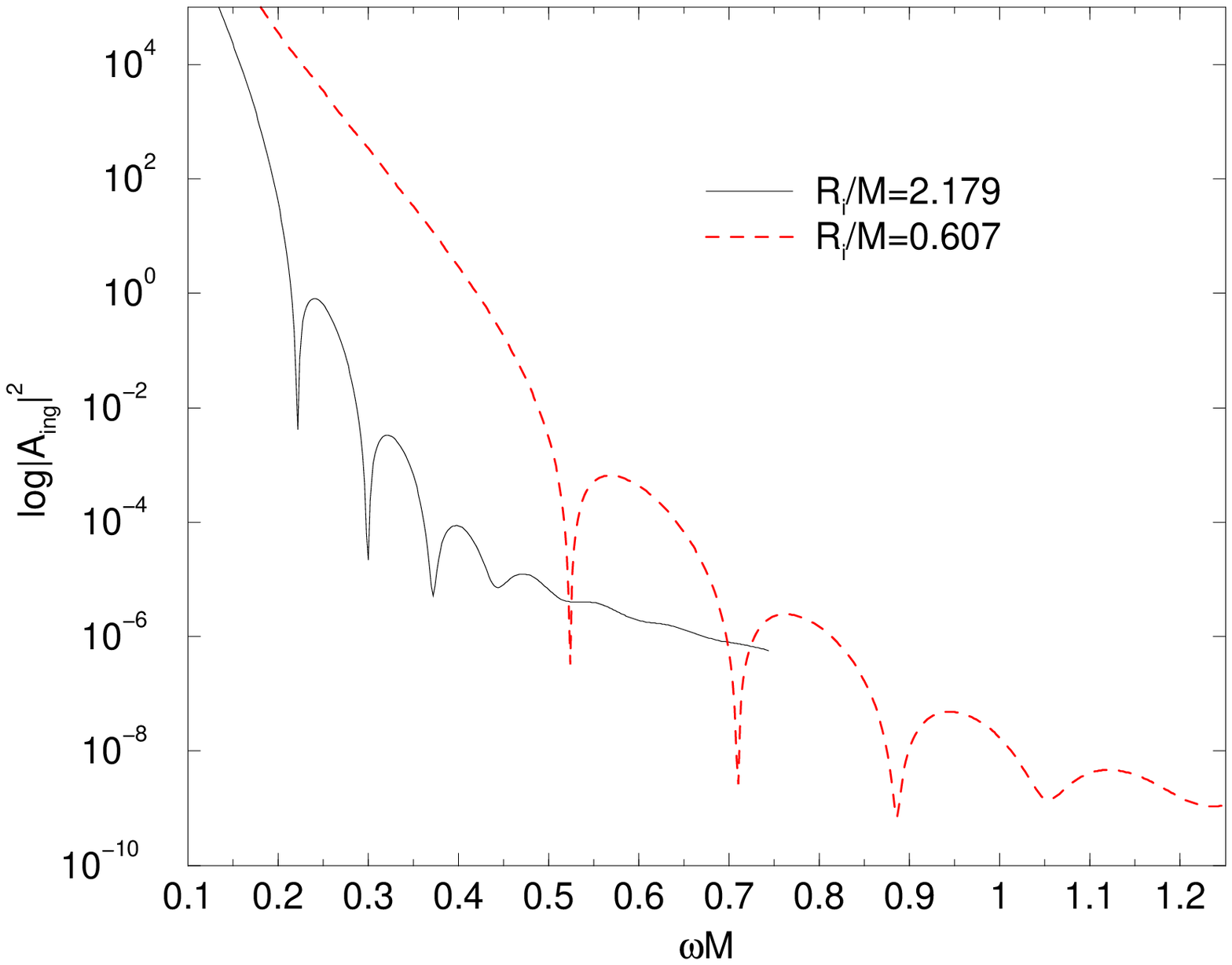}
\caption{Figure 3a displays the odd parity potential for $\ell=2$ and three stellar models: a single density model with $R/M=R_{i}/M_{i}=2.3$ and two double density models both with $\gamma=200, R/M=5, R_{i}/M_{i}=2.3$ but $R_{i}/M=0.607$ and $R_{i}/M=2.179$. Figure 3b shows the least damped $w$-modes of the two DDS stars. The four least QN modes of the star  $R_{i}/M=0.607$ are $\omega_{I}M=0.524+3\times 10^{-6}i$, $\omega_{II}M=0.709+18\times 10^{-5}$, $\omega_{III}M=0.886+3\times 10^{-3}i$ and $\omega_{IV}M=1.054+2\times 10^{-2}i$. The frequencies of the other star are indicated in the text.}
\end{figure}
\begin{figure}
\hspace*{.1\textwidth}
\epsfxsize=0.35\textwidth
\epsfbox{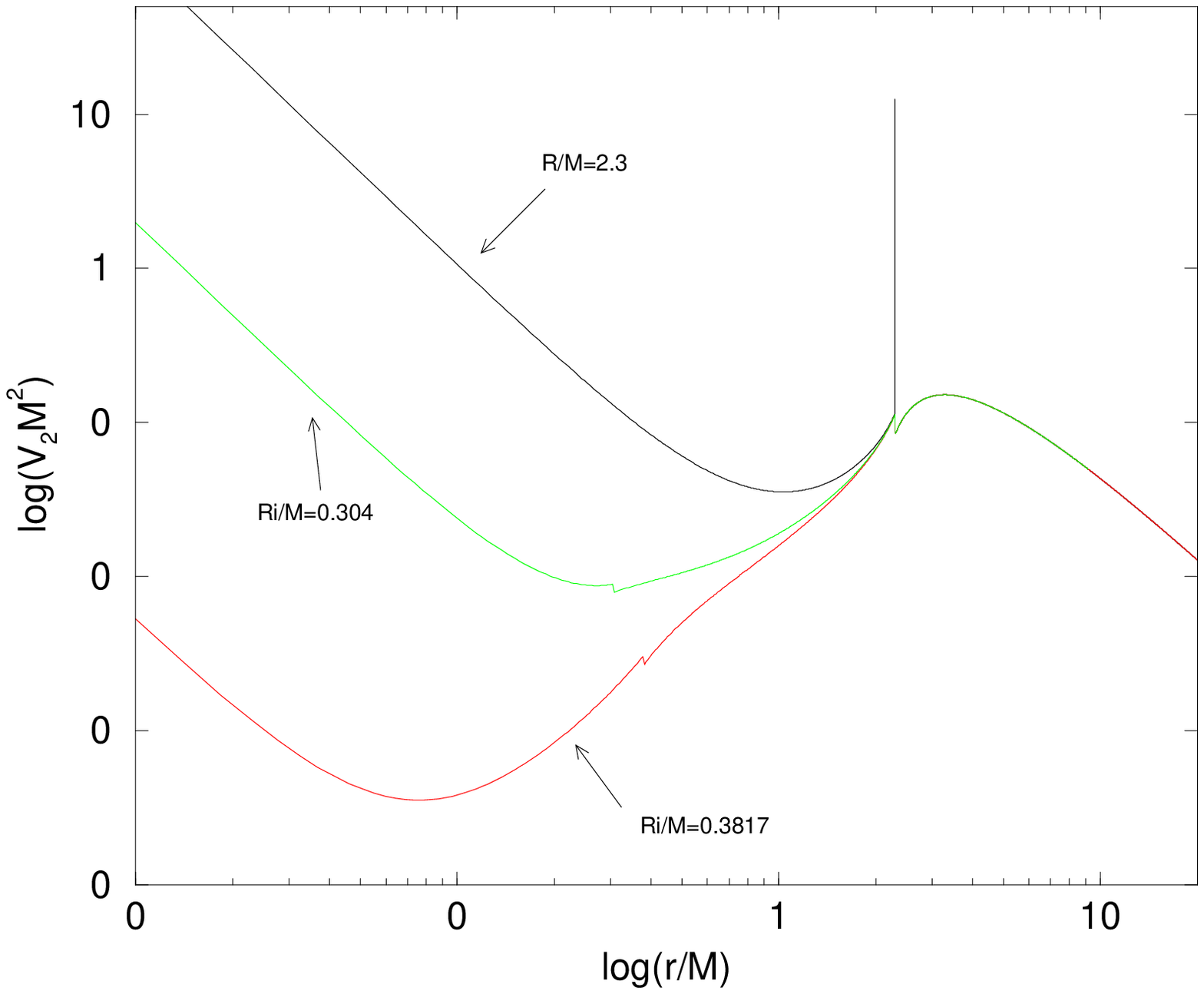}\hspace*{.1\textwidth}\epsfxsize=0.35\textwidth
\epsfbox{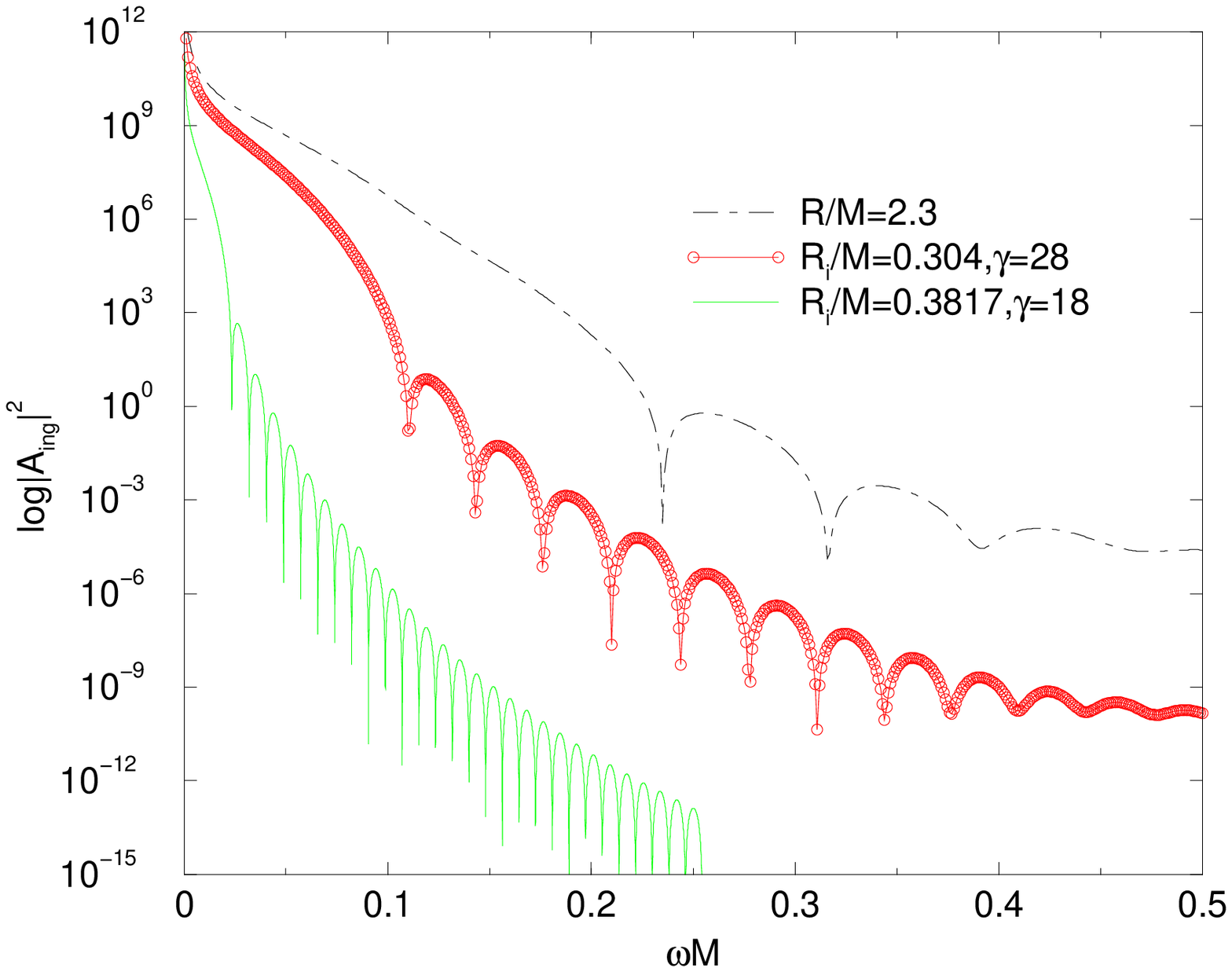}
\caption{Figure 4a displays the $\ell=2$ odd parity potential for three stellar models: A single density star of $R/M=2.3$, a double density star with $\gamma=18, R/M=2.3, R_{i}/M_{i}=5, R_{i}/M=0.3817$ and a double density star with 
$\gamma=28, R/M=2.3, R_{i}/M_{i}=5, R_{i}/M=0.304$. The real part of the least damped quasi-normal modes of these three stellar model are the minima in the graphics of figure 4b. The double density stars are identified, in figure 4a, by their $R_{i}/M$ value and the single density star by its $R/M$ value.}
\end{figure}

\begin{figure}
\hspace*{.1\textwidth}
\epsfxsize=0.35\textwidth
\epsfbox{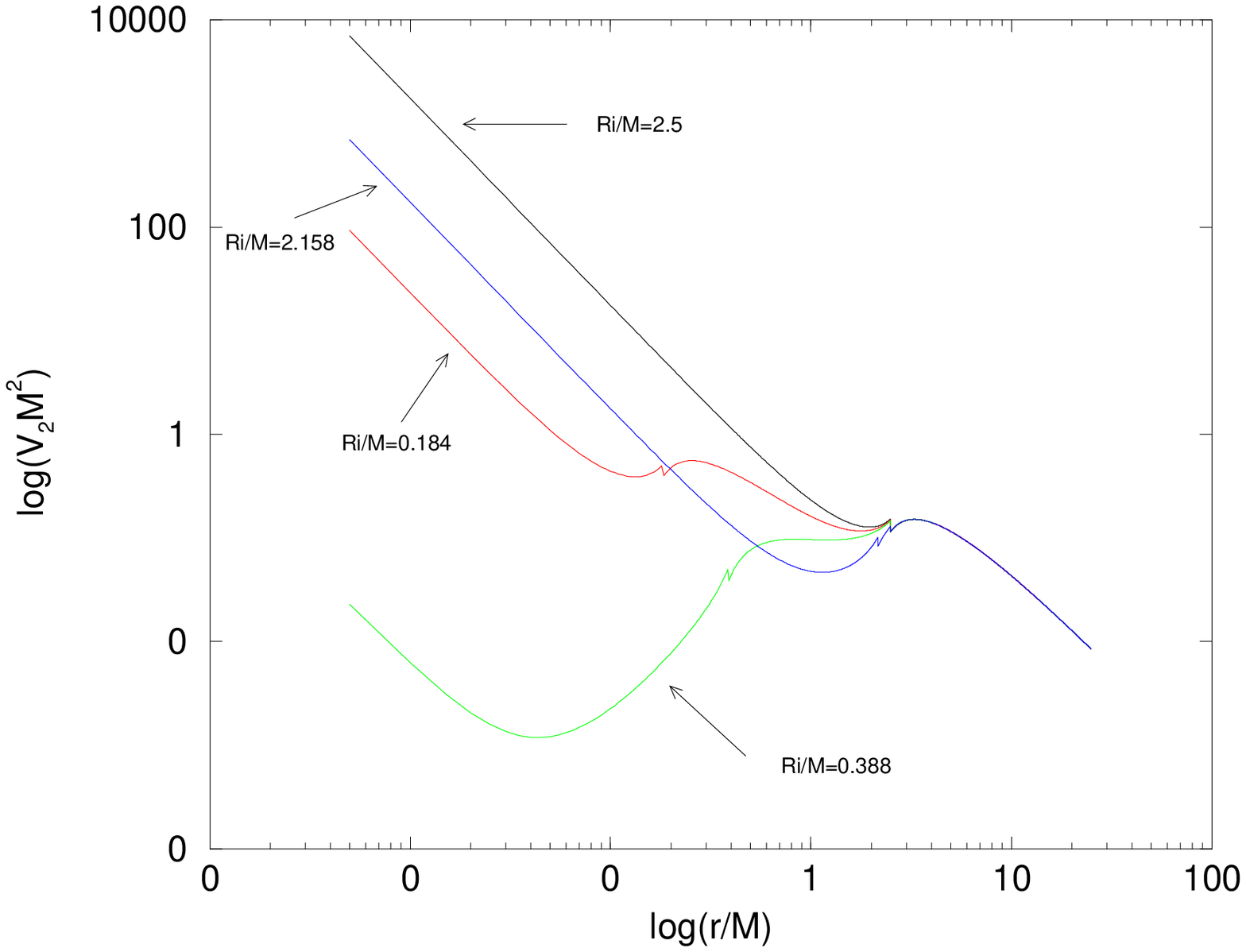}\hspace*{.1\textwidth}\epsfxsize=0.35\textwidth
\epsfbox{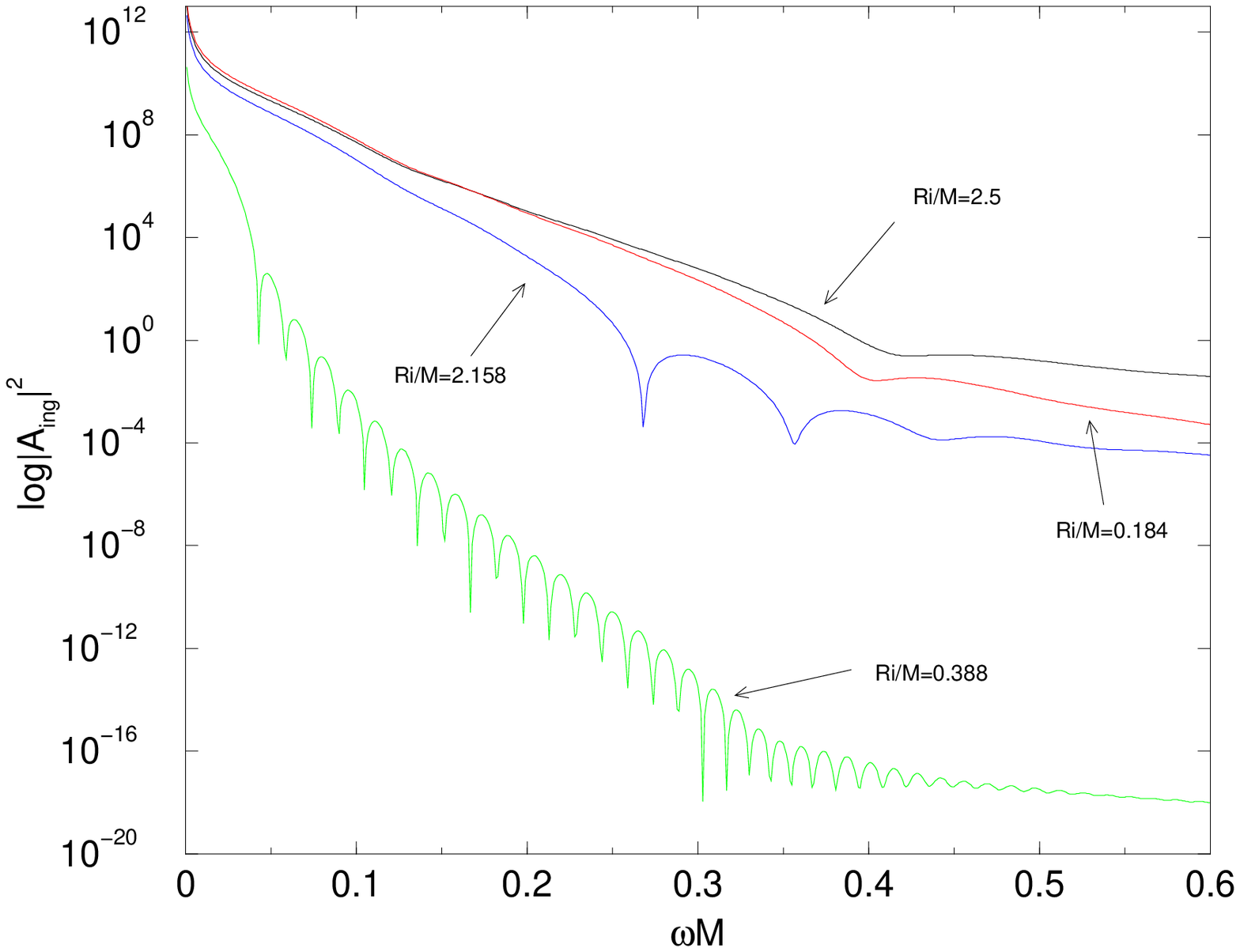}
\caption{Figure 5a displays the $\ell=2$ odd parity potential for four stellar models, all with the same $R/M=R_{i}/M_{i}=2.5$, but different values of $\gamma$: A single density star with $R_{i}/M=2.5$, and three double density stars with $\gamma=3.5, R_{i}/M=2.158$, $\gamma=49, R_{i}/M=0.388$ and $\gamma=200, R_{i}/M=0.184$  As noted in the text, for the three models with $\gamma>3$, there is always the solution corresponding to $R_{i}/M=2.5$, in addition to the solution displayed here. Figure 5b shows the least damped $w$-modes for each of the four models.}
\end{figure}

\begin{figure}
\hspace*{.1\textwidth}
\epsfxsize=0.35\textwidth
\epsfbox{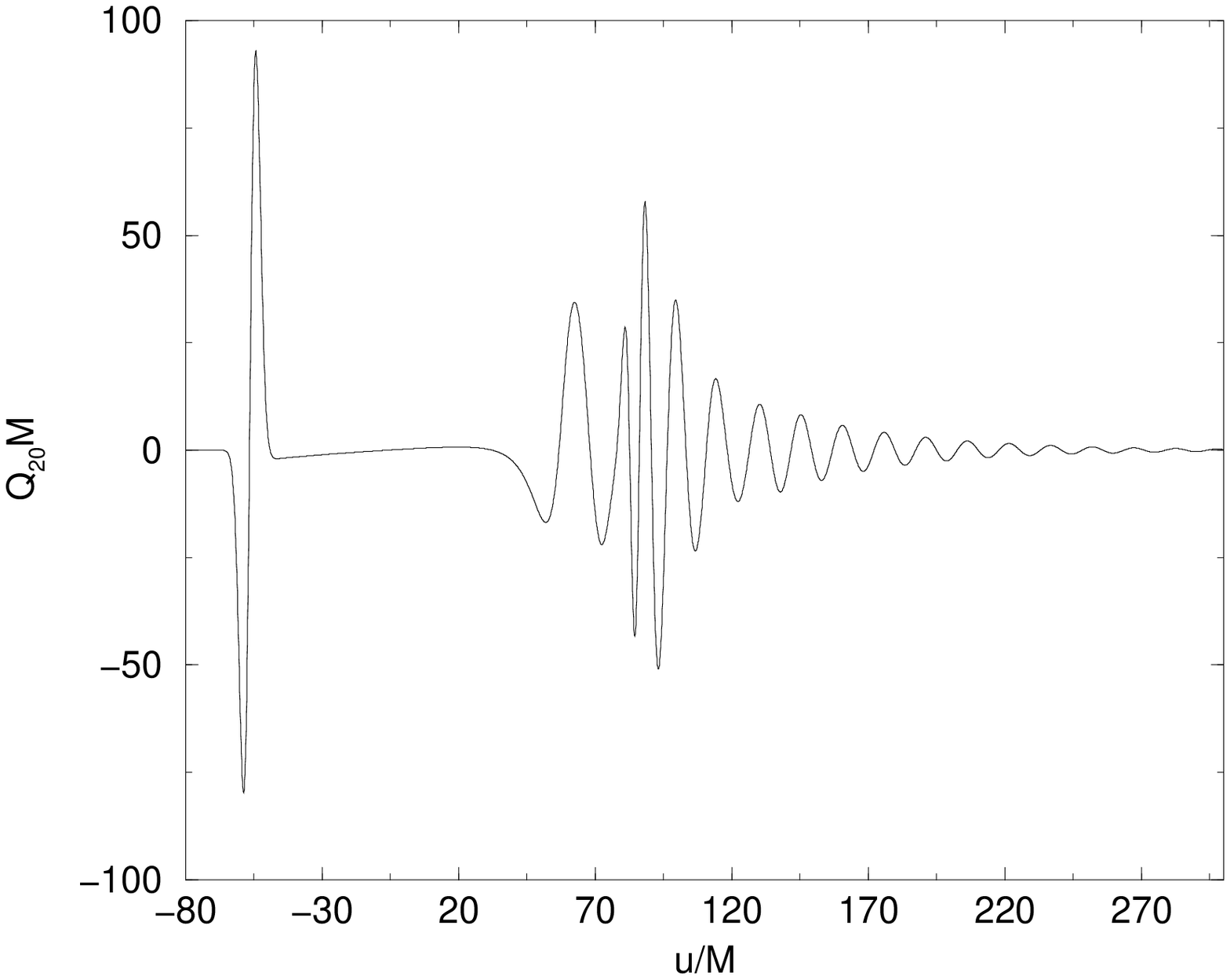}\hspace*{.1\textwidth}\epsfxsize=0.35\textwidth
\epsfbox{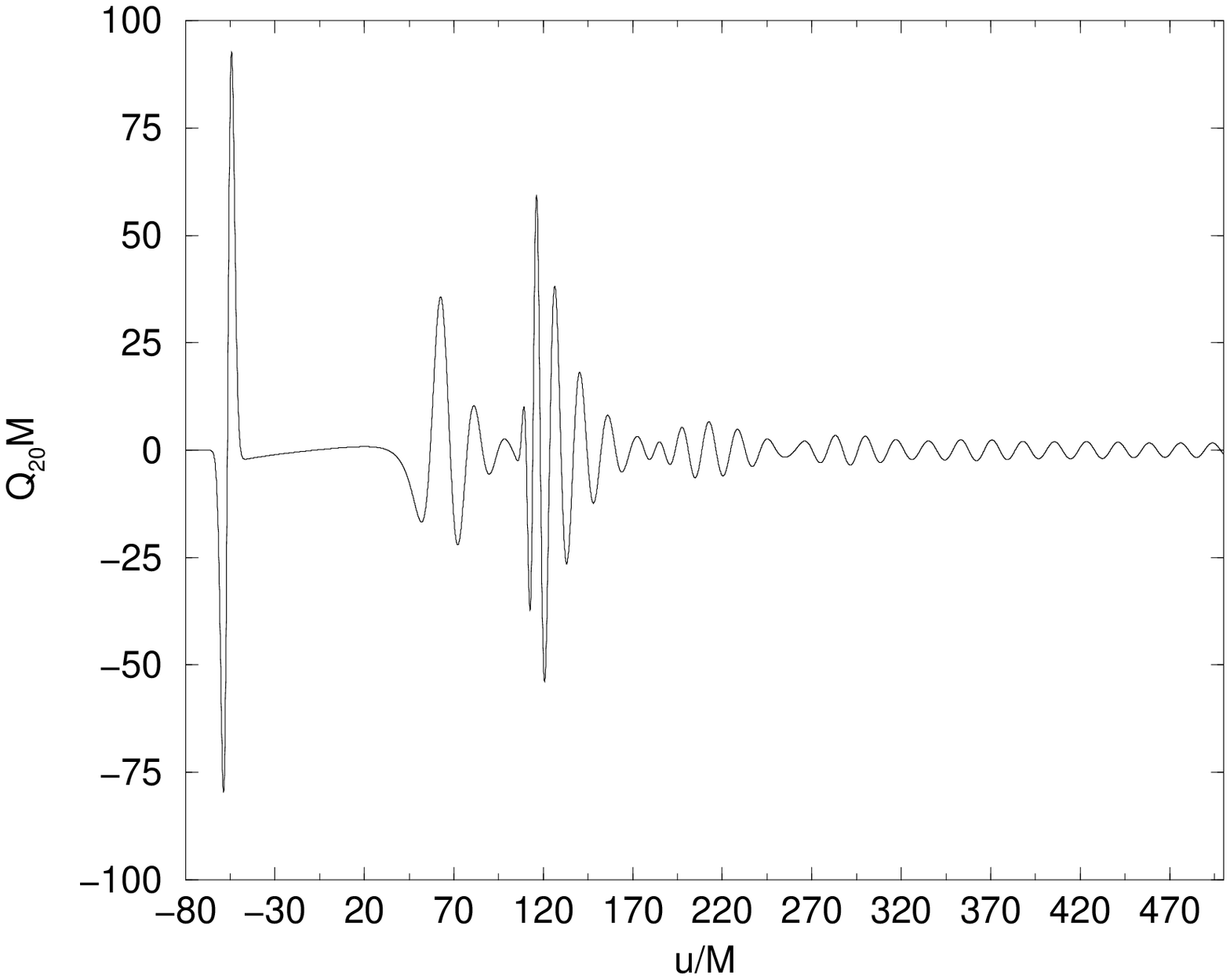}
\caption{Quadrupole ($\ell=2$) waveforms for two stars with $R/M=2.5$ perturbed by matter in a shell located at $R_{shell}=50M$ and with a Gaussian parameter $aM^{2}=0.1$. In figure 6a, a homogeneous star is considered. In figure 6b, a DDS with $R_{i}/M_{i}=2.5,\gamma=3.5$ and $R_{i}/M=2.158$.}
\end{figure}

\begin{figure}
\hspace*{.1\textwidth}
\epsfxsize=0.35\textwidth
\epsfbox{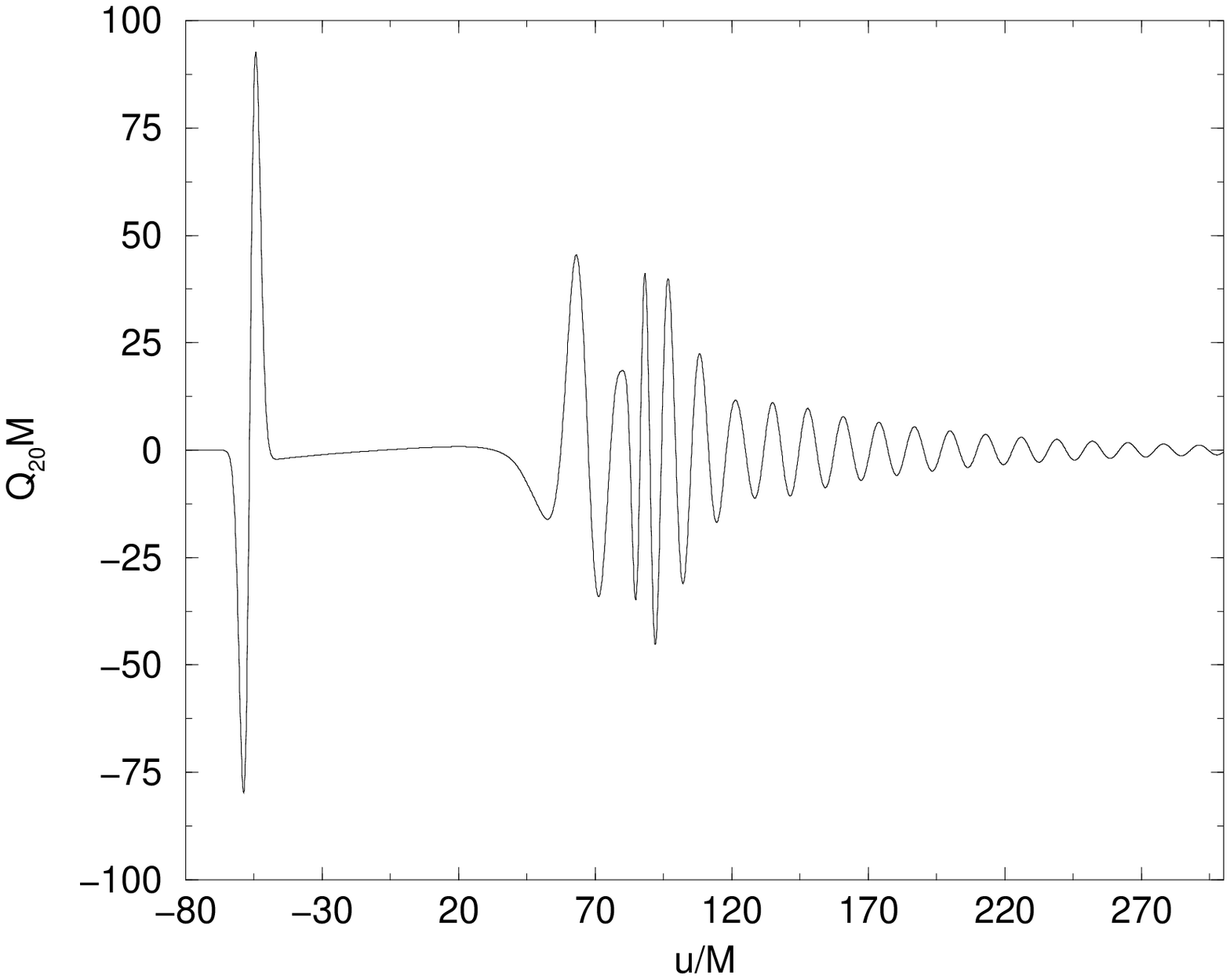}\hspace*{.1\textwidth}\epsfxsize=0.35\textwidth
\epsfbox{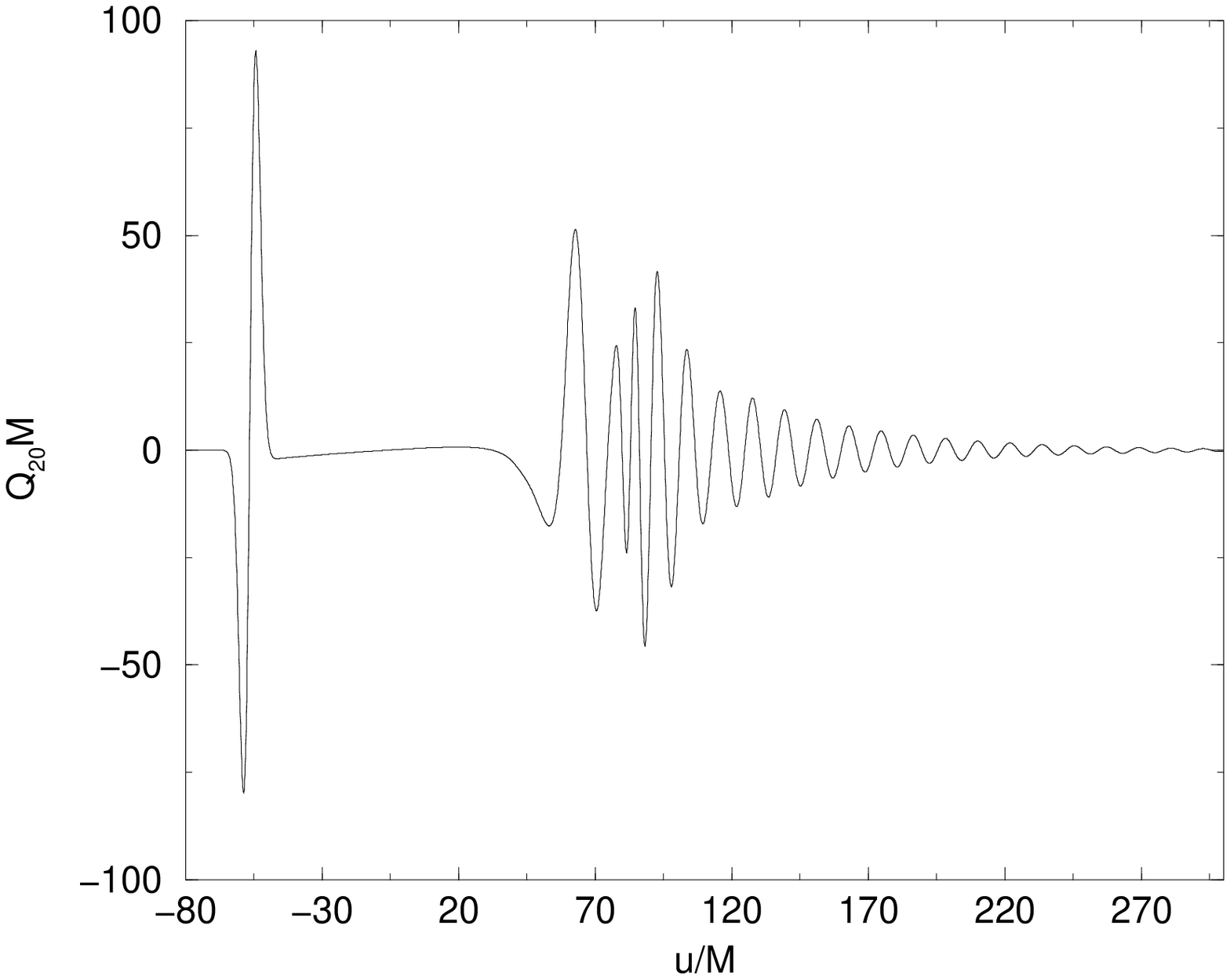}
\caption{Quadrupole ($\ell=2$) waveforms for two stars with $R_{i}/M_{i}=2.5$, perturbed by matter in a shell located at $R_{shell}=50M$ and with a Gaussian parameter $aM^{2}=0.1$. In figure 7a, a DDS with $R/M=5,\gamma=52, R_{i}/M=1.67$. In figure 7b, a DDS with $R/M=6,\gamma=92$ and $R_{i}/M=1.598$.}
\end{figure}

\begin{figure}
\hspace*{.1\textwidth}
\epsfxsize=0.35\textwidth
\epsfbox{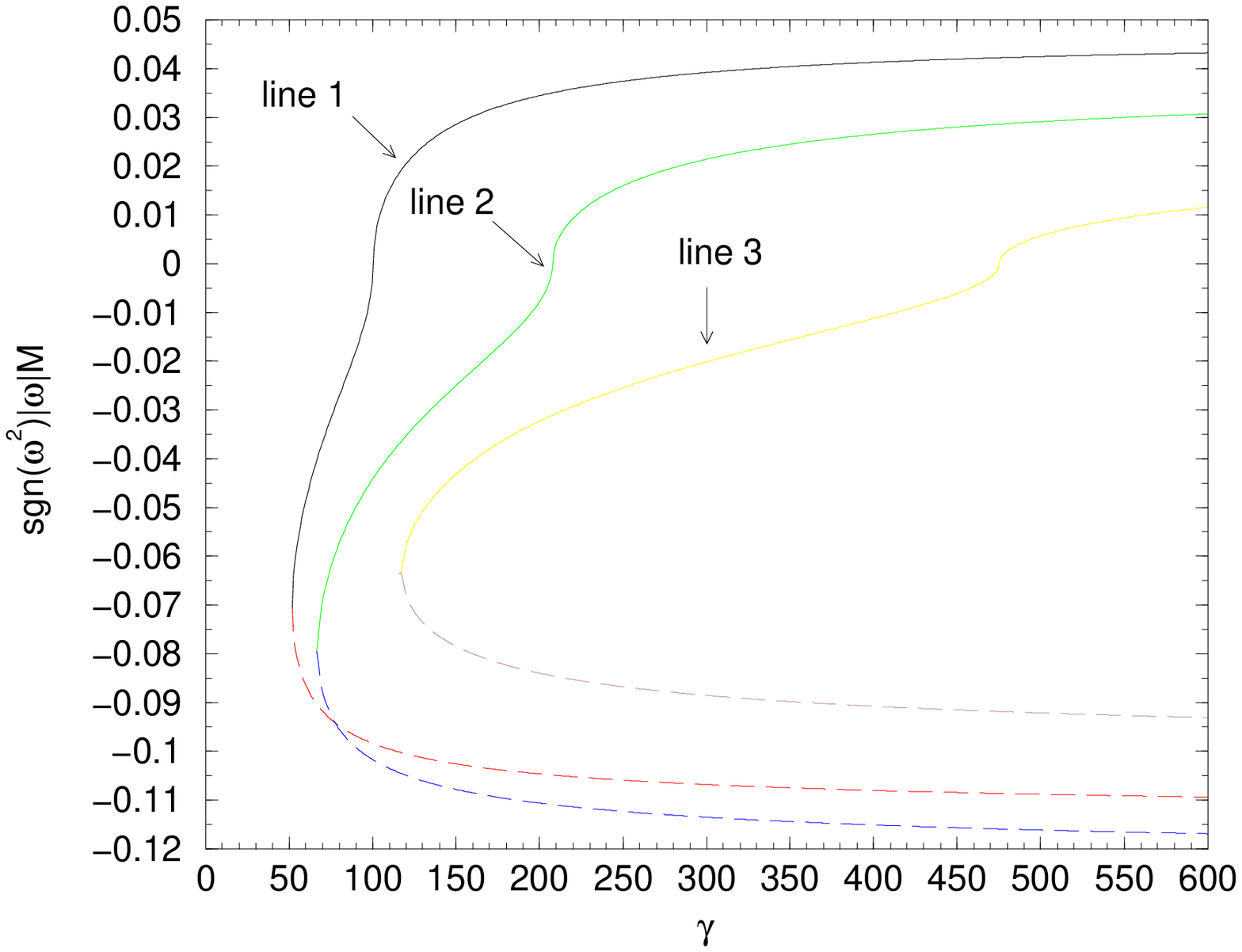}\hspace*{.1\textwidth}\epsfxsize=0.35\textwidth
\epsfbox{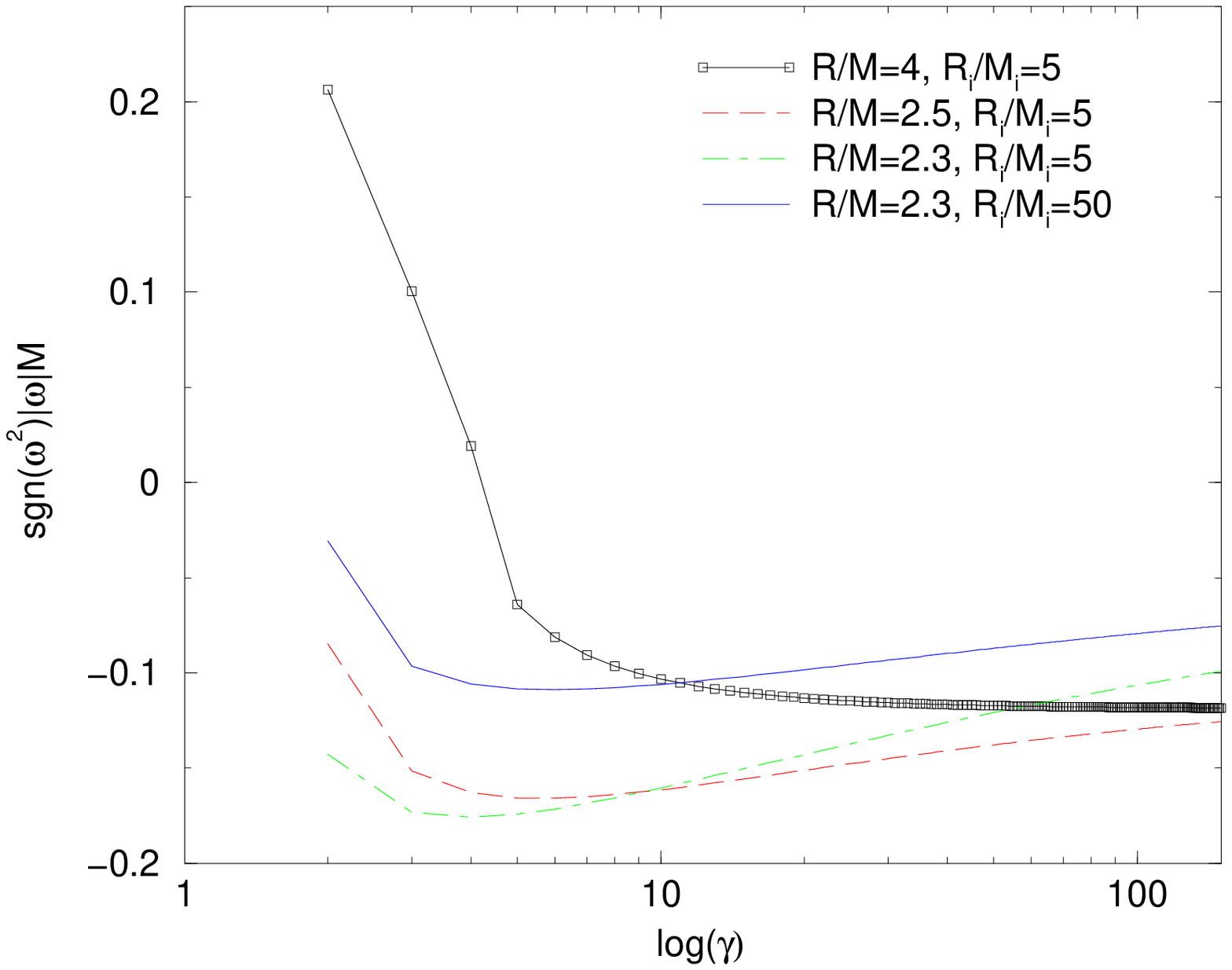}
\caption{Figure 8a displays the radial frequencies of oscillation of a two-phase star. The arrow labeled ``line 1'' represents the frequencies of DDS with $R/M=5, R_{i}/M_{i}=2.5$. The upper curve (in solid) is the solution that approaches $R_{i}/M=2.5$. The lower curve (in dashed) is the solution $R_{i}/M$ that approaches $0$. The arrow labeled ``line 2'' represents the frequencies of DDS with $R/M=5, R_{i}/M_{i}=2.3$. The upper curve (in solid) is the solution that approaches $R_{i}/M=2.3$. The lower curve (in dashed) is the solution $R_{i}/M$ that approaches $0$. The arrow labeled ``line 3'' represents the frequencies of DDS with $R/M=6, R_{i}/M_{i}=2.3$. The upper curve (in solid) is the solution that approaches $R_{i}/M=2.3$. The lower curve (in dashed) is the solution $R_{i}/M$ that approaches $0$. Figure 8b considers models with fixed values of $R/M$ and $R_{i}/M_{i}>R/M$, but varying jump $\gamma$ in the density. All the models are unstable for $\gamma>5$ but  there is a small region near $\gamma\approx 1$ of stable models.}
\end{figure}
\end{document}